\documentclass[reqno,11pt]{amsart}
\usepackage{graphicx}
\usepackage{slashed}
\usepackage{amscd}
\usepackage{amssymb}
\usepackage{esint}
\usepackage[mathscr]{eucal}
\numberwithin{equation}{section}
\allowdisplaybreaks[1]

\title[The Fermionic Projector in an External Potential]{The Fermionic Projector
in a Time-Dependent External Potential: Mass Oscillation Property and Hadamard States}

\author[F.\ Finster, S.\ Murro]{Felix Finster, Simone Murro}
\author[C.\ R\"oken]{Christian R\"oken \\ \\ May 2016}
\thanks{F.F.\ and C.R.\ are supported by the DFG research grant ``Dirac Waves in the Kerr Geometry: Integral Representations, Mass Oscillation Property and the Hawking Effect.''
S.M. is supported within the DFG research training group GRK 1692 ``Curvature, Cycles, and Cohomology.'' }
\address{Fakult\"at f\"ur Mathematik \\ Universit\"at Regensburg \\ D-93040 Regensburg \\ Germany}
\email{finster@ur.de, simone.murro@ur.de, \newline
\hspace*{2.6cm} Christian.Roeken@mathematik.ur.de}

\newtheorem{Def}{Definition}[section]
\newtheorem{Thm}[Def]{Theorem}
\newtheorem{Prp}[Def]{Proposition}
\newtheorem{Lemma}[Def]{Lemma}
\newtheorem{Remark}[Def]{Remark}
\newtheorem{Corollary}[Def]{Corollary}

\newcommand{\Thanks}{\vspace*{.5em} \noindent \thanks}
\newcommand{\beq}{\begin{equation}}
\newcommand{\eeq}{\end{equation}}
\newcommand{\Proof}{\begin{proof}}
\newcommand{\QED}{\end{proof} \noindent}
\newcommand{\QEDrem}{\ \hfill $\Diamond$}

\newcommand{\la}{\langle}
\newcommand{\ra}{\rangle}
\newcommand{\bra}{\mathopen{<}}
\newcommand{\ket}{\mathclose{>}}
\newcommand{\Sl}{\mbox{$\prec \!\!$ \nolinebreak}}
\newcommand{\Sr}{\mbox{\nolinebreak $\succ$}}

\newcommand{\C}{\mathbb{C}}
\newcommand{\R}{\mathbb{R}}
\newcommand{\1}{\mbox{\rm 1 \hspace{-1.05 em} 1}}

\newcommand{\N}{\mathbb{N}}
\newcommand{\Pdd}{\mbox{$\partial$ \hspace{-1.2 em} $/$}}

\newcommand{\V}{\mathscr{V}}

\newcommand\B{{\mathscr{B}}}

\newcommand{\Cisc}{C^\infty_{\text{\rm{sc}}}}
\newcommand{\Cisco}{C^\infty_{\text{\rm{sc}},0}}
\newcommand{\Dir}{{\mathcal{D}}}

\renewcommand{\H}{\mathscr{H}}

\newcommand{\Lin}{\text{\rm{L}}}

\newcommand{\frakD}{\mathfrak{D}}

\newcommand{\p}{\mathfrak{p}}
\newcommand{\SD}{\tilde{\Sig}^\text{\tiny{\rm{D}}}}

\newcommand{\Sig}{\mathscr{S}}

\newcommand{\scrM}{\mycal M}
\newcommand{\scrN}{\mycal N}

\DeclareFontFamily{OT1}{rsfso}{}
\DeclareFontShape{OT1}{rsfso}{m}{n}{ <-7> rsfso5 <7-10> rsfso7 <10-> rsfso10}{}
\DeclareMathAlphabet{\mycal}{OT1}{rsfso}{m}{n}

\begin{document}

\maketitle 

\begin{abstract}
We give a non-perturbative construction of the fermionic projector in Minkowski space
coupled to a time-dependent external potential which is smooth and
decays faster than quadratically for large times.
The weak and strong mass oscillation properties are proven.
We show that the integral kernel of the fermionic projector is of Hadamard form,
provided that the time integral of the spatial sup-norm of the potential satisfies a suitable bound.
This gives rise to an algebraic quantum field theory of Dirac fields in an external potential
with a distinguished pure quasi-free Hadamard state.
\end{abstract}

\tableofcontents

\newpage

\section{Introduction}
Hadamard states play an important role in quantum field theory because
they are needed as a starting point for the perturbative treatment.
Namely, interactions can be dealt with within the framework
of perturbation theory only after a suitable regularization, yielding the so-called
Wick polynomials. By using Hadamard states, one can devise a specific
regularization scheme which is covariant and, thus, especially suited to be applicable
also on curved backgrounds (see for example~\cite{fewster2013necessity, dappiaggiDirac}
or the recent text book~\cite{rejzner}).

While the existence of Hadamard states can be shown abstractly with glueing constructions
(see~\cite{fulling+sweeny+wald, fulling+narcowich+wald}),
these methods are not explicit. Therefore, the construction of Hadamard states remains an important
task. In recent years, two methods for constructing Hadamard states have been developed:
the pseudo-differential operator approach by G{\'e}rard and Wrochna~\cite{gerard2014construction,
gerard+wrochna2, gerard+wrochna3} and the holographic method by Dappiaggi, Moretti and
Pinamonti~\cite{dappiaggi+moretti, dappiaggi+hack+pinamonti, dappiaggi+siemssen, benini+dappiaggi+murro}.
The first method gives a whole class of states.
The second method does give a distinguished state, but it has the shortcoming that it relies on
conformal methods which only apply to massless fields.

Here we propose a new method for constructing Hadamard states for {\em{massive}}
Dirac fields, which gives rise to a distinguished Hadamard state.
The method is based on the recent functional analytic construction in~\cite{finite, infinite} of the
fermionic projector in globally hyperbolic space-times.
In the present paper, we show that the integral kernel of the fermionic projector yields
a pure quasi-free Hadamard state.
In order to make an explicit analysis feasible, we restrict attention to the setting of a time-dependent
external potential in Minkowski space.
We consider this work as a starting point for more general constructions in curved space-times,
to which our methods should also apply.
We point out that our method does not require any symmetries of space-times.
However, if there are space-time symmetries, our state has these symmetries as well.

More specifically, we show that the construction in infinite lifetime in~\cite{infinite} applies to the Dirac equation
in Minkowski space in the presence of an external potential, provided that the potential is smooth and decays suitably for large times.
The main technical step is to prove that the so-called {\em{mass oscillation property}} holds.
Assuming in addition a bound on the time integral of the spatial sup-norm of the potential,
we show that the resulting fermionic projector is of Hadamard form
(for an introduction to the Hadamard form see~\cite{hadamardoriginal, waldQFT}).
These results put the previous perturbative treatment of the fermionic projector
in~\cite{sea, firstorder, light, grotz, norm} (see also the textbook~\cite{PFP}) on a rigorous functional analytic basis.
In particular, our results show that the nonlocal low and high energy contributions as
introduced in~\cite{light} by a formal power series are indeed well-defined and smooth.

In the remainder of the introduction, we state our results and put them into the
context of the fermionic projector and of algebraic quantum field theory.

\subsubsection*{\bf{The Mass Oscillation Property}}
In Minkowski space without an external potential, the Dirac equation has plane-wave solutions.
The sign of the frequency of these plane-wave solutions gives a splitting of the solution space
into two subspaces, usually referred to as the positive and negative energy
subspaces. This frequency splitting is important for the physical interpretation of the Dirac equation and
for the construction of a corresponding quantum field theory.
Namely, choosing the vacuum state in agreement with the frequency
splitting (Dirac sea vacuum), it is possible to reinterpret the negative-energy solutions in terms
of anti-particle states. The plane-wave solutions of positive and negative frequencies
are then identified with creation and annihilation operators, respectively, 
which by acting on the vacuum state generate the whole Fock space.

The above frequency splitting can still be used in static space-times
(i.e.\ if a timelike Killing field is present).
However, in {\em{generic space-times}} or in the presence of a {\em{time-dependent}} external potential,
one does not have a natural frequency splitting. A common interpretation of this fact is that
there is no distinguished ground state, and that the notion particles and anti-particles depend
on the observer. Nonetheless, the construction of the {\em{fermionic projector}} as carried out non-perturbatively
in~\cite{finite, infinite} does give rise to a canonical splitting of the solution space of the Dirac equation
into two subspaces even in generic space-times.
This also suggests that, mimicking the construction for the usual frequency splitting,
there should be a canonical ground state of the
corresponding quantum field theory, even without assuming a Killing symmetry.
One of the goals of this paper is to construct this distinguished
ground state in the presence of a time-dependent external potential in Minkowski space.

We now recall a few basic constructions and definitions from~\cite{infinite}, always restricting attention to
subsets of Minkowski space and to the Dirac equation
\beq \label{Dirext}
\big(i \Pdd + \B(t,\vec{x}) - m \big) \psi_m(t,\vec{x}) =0
\eeq
in the presence of a smooth external potential~$\B$.
Here~$m$ is the rest mass, and for clarity we add it as an index to the wave function.
The construction differs considerably in the cases when the space-time
has finite or infinite lifetime.
A typical example of a space-time of {\em{finite lifetime}} is an open
subset~$\Omega$ of Minkowski space contained in a strip~$(-T,T) \times \R^3$ for some constant~$T>0$
such that the surface~$\{0\} \times \R^3$ is a Cauchy surface (for a general treatment of
space-times of finite lifetime see~\cite{finite}). In this case, one considers on the
solution space of the Dirac equation~\eqref{Dirext}
the usual scalar product obtained by integrating over the Cauchy surface\footnote{The factor~$2 \pi$ might seem
unconventional. This convention was first adopted in~\cite{rrev} to simplify some formulas.}
\beq \label{printintro}
(\psi_m | \phi_m)_m = 2 \pi \int_{\R^3} \Sl \psi_m | \gamma^0 \phi_m \Sr|_{(t=0, \vec{x})}\: d^3x
\eeq
as well as the space-time inner product
\beq \label{stipintro}
\bra \psi_m | \phi_m \ket = \int_\Omega \Sl \psi_m | \phi_m \Sr_x \: d^4x
\eeq
(here~$\Sl \psi | \phi \Sr$ is the spin scalar product, which is often denoted by~$\overline{\psi} \phi$
with the adjoint spinor~$\overline{\psi} = \psi^\dagger \gamma^0$, where the dagger means
complex conjugation and transposition).
Finite lifetime implies that the space-time inner product is bounded in the sense that
there is a constant~$c>0$ such that
\beq \label{finitebound}
|\bra \phi_m | \psi_m \ket| \leq c \:\|\phi_m\|_m\: \|\psi_m\|_m
\eeq
for all Dirac solutions~$\psi_m, \phi_m$ (and~$\|.\|_m$ is the norm corresponding to
the scalar product~\eqref{printintro}). This in turn makes it possible to
represent the space-time inner product in terms of the {\em{fermionic signature operator}}~$\tilde{\Sig}$,
meaning that there is a unique bounded symmetric operator~$\tilde{\Sig}$ such that
\[ \bra \phi_m | \psi_m \ket = ( \phi_m \,|\, \tilde{\Sig}\, \psi_m)_m \]
(here the tilde indicates that an external potential~$\B$ is present, whereas the corresponding
objects in the Minkowski vacuum are denoted without a tilde).
Then the positive and negative spectral subspaces
of~$\tilde{\Sig}$ give rise to the desired splitting of the solution space.

The above construction fails in space-times of {\em{infinite lifetime}} because the
time integral in~\eqref{stipintro} will in general diverge. The way out is to consider
families of solutions~$(\psi_m)_{m \in I}$ of the family of Dirac equations~\eqref{Dirext} with
the mass parameter~$m$ varying in an open interval~$I$.
We need to assume that~$I$ does not contain the origin, because our methods 
for dealing with infinite lifetime do not
apply in the massless case~$m=0$ (this seems no physical restriction because all known
fermions in nature have a non-zero rest mass). By symmetry, it suffices to consider positive masses.
Thus we choose
\beq \label{Idef}
I:=(m_L, m_R) \subset \R \qquad \text{with parameters~$m_L, m_R>0$}\:.
\eeq
We always choose the family of solutions~$(\psi_m)_{m \in I}$ in the
class~$\Cisco(\scrM \times I, S\scrM)$ of smooth solutions with spatially compact support
in Minkowski space~$\scrM$ which depend smoothly on~$m$ and vanish identically for~$m$ outside a
compact subset of~$I$. On such families of solutions, we can impose conditions
analogous to~\eqref{finitebound} by suitably integrating over~$m$.
We here give the condition which is most relevant for applications
(for a weaker version, which will also arise in intermediate steps of our proofs,
see Definition~\ref{defWMass Oscillation Property} below).

\begin{Def}
The Dirac operator~$i \Pdd + \B$ has the {\bf{strong mass oscillation property}} in the
interval~$I$ (see~\eqref{Idef}) if there is a constant~$c>0$ such that
\beq \label{smopintro}
\bigg| \bra \int_I \phi_m \,dm \:|\: \int_I \psi_{m'} \,dm'  \ket \bigg|
\leq c \int_I \, \|\phi_m\|_m\, \|\psi_m\|_m\: dm
\eeq
for all families of solutions~$(\psi_m)_{m \in I},  (\phi_m)_{m \in I} \in \Cisco(\scrM \times I, S\scrM)$.
\end{Def} \noindent
The point is that we integrate over the mass parameter {\em{before}} taking the space-time inner product.
Intuitively speaking, integrating over the mass parameter generates
a decay of the wave function, making sure that the time integral converges.

As shown in~\cite[Section~4]{infinite}, the strong mass oscillation property
gives rise to the representation
\[ \bra \int_I \phi_m \,dm \,|\, \int_I \psi_{m'} \,dm' \ket = \int_I (\phi_m \,|\, \tilde{\Sig}_m \,\psi_m)_m\: dm \:, \]
which for every~$m \in I$ uniquely defines the {\em{fermionic signature operator}}~$\tilde{\Sig}_m$.
This operator is bounded and symmetric with respect to the scalar product~\eqref{printintro}.
Moreover, it does not depend on the choice of the interval~$I$.
Now the positive and negative spectral subspaces
of the operator~$\tilde{\Sig}_m$ again yield the desired splitting of the solution space.

The remaining crucial question is whether the inequality~\eqref{smopintro} holds
in the presence of an external potential. In the case of a {\em{static potential}},
it is shown in~\cite[Theorem~5.1]{infinite} that the mass oscillation property holds,
and that the fermionic projector gives back the usual frequency splitting.
In particular, if~$\B$ vanishes, the fermionic projector gives the usual vacuum ground state.
With these results in mind, we here consider the {\em{time-dependent situation}}.
Also in this setting, we give an affirmative answer to the above question,
provided that the potential has suitable decay properties at infinity:
\begin{Thm} \label{thmmop}
Assume that the external potential~$\B$ is smooth and for large times
decays faster than quadratically in the sense that
\beq |\B(t)|_{C^2} \leq \frac{c}{1 + |t|^{2+\varepsilon}} \label{Bdecay}
\eeq
for suitable constants~$\varepsilon, c>0$. Then the strong mass oscillation property holds.
\end{Thm} \noindent
The $C^2$-norm in~\eqref{Bdecay} is defined as follows.
We denote spatial derivatives by~$\nabla$ and use the
notation with multi-indices, i.e.\ for a multi-index $\alpha = (\alpha_1, \ldots, \alpha_p)$
we set~$\nabla^\alpha = \partial_{\alpha_1 \cdots \alpha_p}$ and
denote the length of the multi-index by~$|\alpha|=p$.
Then the spatial $C^k$-norms of the potential are defined by
\beq \label{defCk}
|\B(t)|_{C^k} := \max_{|\alpha| \leq k} \;\sup_{\vec{x} \in \R^3} |\nabla^\alpha \B(t, \vec{x})| \:,
\eeq
where~$| \,.\, |$ is the $\sup$-norm corresponding to the norm~$|\phi|^2 := \Sl \phi | \gamma^0 \phi \Sr$
on the spinors.

\subsubsection*{\bf{The Fermionic Projector and the Hadamard Form}}
Using the result of the previous theorem,
the fermionic projector~$P$ is defined for a fixed mass parameter~$m \in I$
by (see~\cite[Definition~3.7]{finite} and~\cite[Definition~4.5]{infinite})
\beq \label{Pdef}
P := -\chi_{(-\infty, 0)}(\tilde{\Sig}_m)\, \tilde{k}_m \:,
\eeq
where~$\chi_{(-\infty, 0)}(\tilde{\Sig}_m)$ is the projection onto the negative spectral subspace of
the fermionic signature operator, and~$\tilde{k}_m$ is the causal fundamental solution
(for basic definitions see Section~\ref{secgreen} below).
The fermionic projector can be represented as a bi-dis\-tri\-bu\-tion~$P(x,y)$
with~$x,y \in \scrM$
(see~\cite[Section~3.5]{finite} and~\cite[Section~4.3]{infinite}), which satisfies the Dirac equation
and is symmetric, i.e.
\begin{align}
(i \Pdd_x + \B(x) - m)\, P(x,y) &= 0 \label{PDirac} \\
P(x,y)^* = P(y,x) \label{Psymm}
\end{align}
(where~$P(x,y)^* = \gamma^0 P(x,y)^\dagger \gamma^0$ is the adjoint with respect to
the spin scalar product~$\Sl .|. \Sr$).

Knowing the singularity structure of the bi-dis\-tri\-bu\-tion~$P(x,y)$
is important for applications (point-splitting method, Wick polynomials, renormalization, etc.).
Therefore, we shall establish that the fermionic projector is of {\em{Hadamard form}}.
In our setting, this is tantamount to proving that the
bi-distribution~$P(x,y)$ is of the form (see~\cite{sahlmann2001microlocal} or~\cite[page~156]{hack})
\beq \label{hadamard1}
P(x,y) = \lim_{\varepsilon \searrow 0} \;i \Pdd_x \left( \frac{U(x,y)}{\sigma_\varepsilon(x,y)}
+ V(x,y)\: \log \sigma_\varepsilon(x,y) + W(x,y) \right) ,
\eeq
where
\beq \label{sigmadef}
\sigma_\varepsilon(x,y) := (y-x)^j\, (y-x)_j - i \varepsilon \,(y-x)^0 \:,
\eeq
and~$U$, $V$ and~$W$ are smooth functions on~$\scrM \times \scrM$
taking values in the~$4 \times 4$-matrices acting on the spinors
(we always denote space-time indices by latin letters running from~$0, \ldots, 3$).
For clarity, we point out that on a manifold, the function~$\sigma_\varepsilon$
can be defined locally in a geodesically convex neighborhood of a point~$x \in \scrM$,
making it necessary to distinguish between the {\em{local}} Hadamard form (i.e.\ a local representation
of the form~\eqref{hadamard1}) and the {\em{global}} Hadamard form (implying that the singularities
in~\eqref{hadamard1} are the only singularities of the bi-distribution).
For these subtle issues, we refer the reader to~\cite{gonnella, fulling+sweeny+wald, radzikowski}.
In our setting of Minkowski space, there is one global
chart, and the distance function~$\sigma_\varepsilon$ is defined globally by~\eqref{sigmadef}.
For this reason, we do not need to make a distinction between the local Hadamard form and the
global Hadamard form.

In space-times of {\em{finite lifetime}}, in general the fermionic projector is {\em{not}} of
Hadamard form. The first counter examples were constructed in~\cite{fewster+lang},
where it is shown that in ultrastatic space-times of finite lifetime,
the Hadamard condition is in general violated. Other counter examples are so-called
simple domains as introduced in~\cite[Sections~2.1 and~1.4]{drum}. In such simple domains,
the fermionic projector in the massless case is an operator of finite rank with a pointwise bounded integral kernel,
clearly not being of Hadamard form.

In space-times of {\em{infinite lifetime}}, the situation is better at least in the ultrastatic case.
Namely, in~\cite[Section~5]{infinite} it is shown that the fermionic projector in ultrastatic space-times
is composed of all negative-energy solutions of the Dirac equation.
Therefore, $P(x,y)$ coincides with the bi-distribution constructed
from the frequency splitting, which in~\cite{wrochna2012quantum} was shown to be
of Hadamard form.
Apart from this specific result, it is unknown whether or in which space-times the 
fermionic projector is of Hadamard form.

The main result of this paper is to show that
in a time-dependent external potential in Minkowski space, the fermionic projector is indeed of Hadamard form,
provided that the potential is not too large:

\begin{Thm} \label{thmHadamard}
Assume that the external potential~$\B$ is smooth, and that its time derivatives decay
at infinity in the sense that~\eqref{Bdecay} holds and in addition that
\[ \int_{-\infty}^\infty |\partial_t^p \B(t)|_{C^0}\, dt < \infty \qquad \text{for all~$p \in \N$} \]
(with the $C^0$-norm as defined in~\eqref{defCk}).
Moreover, assume that the potential satisfies the bound
\beq
\int_{-\infty}^\infty |\B(t)|_{C^0}\, dt < \sqrt{2}-1 \:. \label{Bssmall}
\eeq
Then the fermionic projector~$P(x,y)$ is of Hadamard form.
\end{Thm}

We note that the property of the bi-distribution~$P(x,y)$ to be of
Hadamard form can also be expressed in terms of the wave front set
(see for example~\cite{hormanderI, strohmaierML}).
Also, the smooth functions in~\eqref{hadamard1} can be expanded in
powers of the Minkowski distance~$\sigma(x,y) = (y-x)^j \, (y-x)_j$,
\[ U(x,y) = \sum_{n=0}^\infty U_n(x,y)\, \sigma^n\:,\quad
V(x,y) = \sum_{n=0}^\infty V_n(x,y)\, \sigma^n\:,\quad
W(x,y) = \sum_{n=0}^\infty W_n(x,y)\, \sigma^n\:. \]
The coefficients of this so-called {\em{Hadamard expansion}}
can be computed iteratively using the method of {\em{integration along characteristics}}
(see~\cite{hadamardoriginal, friedlander1} or~\cite{baer+ginoux}).
In Minkowski space, the {\em{light-cone expansion}}~\cite{firstorder, light} gives a systematic
procedure for computing an infinite number of Hadamard coefficients in one step.
This procedure also makes it possible to compute the {\em{smooth
contributions}} to~$P(x,y)$, giving a connection to fermionic loop corrections in quantum field theory
(see~\cite[\S8.2 and Appendix~D]{sector}).

We finally remark that, introducing an ultraviolet regularization, the fermionic projector
gives rise to a corresponding {\em{causal fermion system}} (for details see~\cite[Section~4]{finite}).
In this context, the result of Theorem~\ref{thmHadamard} gives a justification for
the formalism of the continuum limit as used in~\cite{PFP, cfs} for the analysis of the causal action principle.
We also refer the interested reader to the introduction to causal fermion systems~\cite{intro}.

\subsubsection*{\bf{Quantum Fields and Hadamard States}}
Using the standard notation in quantum field theory,
the objective of the quantization of the Dirac field is to construct field operators~$\Psi(x)$
and~$\Psi(y)^*$ acting on a Fock space~$\H_\text{\tiny{Fock}}$ together with a suitable
ground state~$|0\ra$.
The field operators should satisfy the canonical anti-commutation
relations
\begin{align}
\{\Psi^\alpha(x),\Psi^\beta(y)^* \} &= \big( \tilde{k}_m(x,y) \big)^\alpha_\beta \label{commnew1} \\
\{\Psi^\alpha(x),\Psi^\beta(y)\} & = \{\Psi^\alpha(x)^*,\Psi^\beta(y)^*\} = 0\:,
\end{align}
where the Greek indices running from~$1,\dots, 4$ denote Dirac spinor indices
(we always work in natural units~$\hbar = c = 1$).
For clarity, we point out that, due to the conjugation of the operator~$\Psi^\beta(y)$ in~\eqref{commnew1},
the spinor index~$\beta$ should be thought of as the index of a dual spinor; this is why it appears on the right side
as lower index.
In the presence of a time-dependent external potential, there is no distinguished ground state.
But there is common agreement that for a physically sensible theory the
ground state~$|0\ra$ should be chosen such that
the two-point function~$\la 0 |\, \Psi(x) \,\Psi(y)^* \,| 0 \ra$ is a bi-distribution
of Ha\-da\-mard form.
In this paper, we shall achieve this goal by arranging that the two-point function
coincides with the bi-distribution of the fermionic projector, i.e.\
\[ \la 0 |\, \Psi^\alpha(x) \,\Psi^\beta(y)^* \,| 0 \ra = - \big(P(x,y)\big)^\alpha_\beta \:. \]

In order to give the above formulas a mathematical meaning, one needs to ``smear out'' the
field operators and work with operator-valued distributions
(see for example~\cite{streater+wightman, dimock1, dimock3, dappiaggiDirac}).
Formally, this is accomplished by setting
\[ \Psi(g) = \int_\scrM \Psi(x)^\alpha\: g(x)_\alpha\: d^4x \qquad \text{and} \qquad
\Psi^*(f) = \int_\scrM \big(\Psi(x)^\alpha\big)^* \: f(x)^\alpha\: d^4x \:, \]
where~$g$ and~$f$ are smooth and compactly supported co-spinors and spinors, respectively.
We do not aim at defining the pointwise field operators,
but instead we work exclusively with the smeared field operators~$\Psi(g)$ and~$\Psi^*(f)$
(for basic definitions see Section~\ref{secphysics}).
Moreover, instead of considering
vacuum expectation values, in the algebraic formulation of quantum field theory one prefers to work with
a pure quasi-free state~$\omega$, making it unnecessary choose a representation of
the field algebra on the Fock space. Given a state~$\omega$, a corresponding representation
of the field algebra is obtained by applying the GNS construction,
also making it possible to recover~$\omega$ as a vacuum expectation value.
A quasi-free state for which the two-point function is of Hadamard form~\eqref{hadamard1}
is called a {\em{Hadamard state}}.

Using the algebraic language, we prove the following result:
\begin{Thm} \label{thmstate}
There is an algebra of smeared fields generated by~$\Psi(g)$, $\Psi^*(f)$
together with a pure quasi-free state~$\omega$ with the following properties: \\[0.3em]
(a) The canonical anti-commutation relations hold:
\beq \label{antismeared}
\{\Psi(g),\Psi^*(f)\} = \bra g^* \,|\, \tilde{k}_m\, f \ket \:,\qquad
\{\Psi(g),\Psi(g')\} = 0 = \{\Psi^*(f),\Psi^*(f')\} \:.
\eeq
(b) The two-point function of the state is given by
\[ \omega \big( \Psi(g) \,\Psi^*(f) \big) = -\iint_{\scrM \times \scrM} g(x) P(x,y) f(y) \: d^4x\, d^4y\:. \]
\end{Thm} \noindent
The main step in the proof is to use the spectral projection operators~$\chi_{(-\infty, 0)}(\tilde{S}_m)$
and $\chi_{[0,\infty)}(\tilde{S}_m)$ to construct a positive operator~$R$,
making it possible to apply Araki's results in~\cite{araki1970quasifree}
to obtain the desired quasi-free state.

We finally put our result into the context of other methods for constructing Hada\-mard states.
First, there is the method of glueing the physical space-time to an ultrastatic space-time and
using that the Hadamard property is preserved under time evolution (see~\cite{fulling+sweeny+wald, fulling+narcowich+wald}). This method shows the existence of Hadamard states in every
globally hyperbolic space-time and gives a constructive procedure for
a class of Hadamard states.
Another method is to work with pseudo-differential operators~\cite{gerard2014construction},
again giving a whole class of Hadamard states.
A method which distinguishes one specific Hadamard state using asymptotic symmetries at null infinity
is given in~\cite{dappiaggi+moretti}.
Our method gives a {\em{unique distinguished Hadamard state}} even in
the generic time-dependent setting in Minkowski space. 
Moreover, this method is constructive in the sense that
the bi-distribution~$P(x,y)$ and its Hadamard expansion can be computed explicitly
(see~\cite{norm, light}). Our results exemplify that 
the construction of the fermionic projector in~\cite{infinite} is a promising method
for constructing a distinguished Hadamard state without
any symmetry assumptions, hopefully even in generic globally hyperbolic space-times.

\section{Preliminaries}
\subsection{Dirac Green's Functions and the Time Evolution Operator} \label{secgreen}
Let~$\scrM$ be Minkowski space, a four-dimensional real vector space endowed with an
inner product of signature~$(+ \ \!\! - \ \!\! - \ \! - )$.
The Dirac equation in the Minkowski vacuum (i.e.\ without external potential) reads
\[ (i \Pdd - m)\, \psi(x) = 0 \;, \]
where we use the slash notation with the Feynman dagger~$\Pdd := \gamma^j \partial_j$.
We always work with the Dirac matrices in the Dirac representation
\[ \gamma^0 = \left( \begin{array}{cc} \1_{\mathbb{C}^2} & 0 \\ 0 & -\1_{\mathbb{C}^2} \end{array} \right) ,\qquad \vec{\gamma} = \left( \begin{array}{cc}
0 & \vec{\sigma} \\ -\vec{\sigma} & 0 \end{array} \right) \]
(and~$\vec{\sigma}$ are the three Pauli matrices).
The wave functions at a space-time point~$x$ take values in the {\em{spinor space}}~$S_x$, a four-dimensional complex vector space endowed with an indefinite scalar product of signature $(2,2)$,
which we call {\em{spin scalar product}} and denote by
\[ \Sl \psi | \phi \Sr_x = \sum_{\alpha=1}^4 s_\alpha\: \psi^\alpha(x)^\dagger
\phi^\alpha(x) \:,\qquad s_1=s_2=1,\;\; s_3=s_4=-1\:, \]
where $\psi^\dagger$ is the complex conjugate wave function (this scalar product is
often written as $\overline{\psi} \phi$ with the so-called adjoint spinor
$\overline{\psi} = \psi^\dagger \gamma^0$).
We denote the space of smooth wave functions by~$C^\infty(\scrM, S\scrM)$,
whereas~$C^\infty_0(\scrM, S\scrM)$ denotes the smooth and compactly supported wave functions (here $S\scrM$ is the spinor bundle over Minkowski space with fibers $S_x \scrM$).
On the spaces of wave functions, one can introduce a Lorentz-invariant pairing by
integrating the spin scalar product over space-time,
\begin{gather}
\bra .|. \ket \::\: C^\infty(\scrM, S\scrM) \times C^\infty_0(\scrM, S\scrM) \rightarrow \C \:, \notag \\
\bra \psi|\phi \ket = \int_\scrM \Sl \psi | \phi \Sr_x \: d^4x\:.  \label{stip} 
\end{gather}

In what follows, the mass parameter of the Dirac equation~$m$ will not be fixed. It can vary in
an open interval~$I:=(m_L, m_R)$ with~$m_L, m_R>0$.
In order to make this dependence explicit, we often add the mass as an index.
Moreover, we consider an {\em{external potential}}~$\B$, which we assume to be
symmetric with respect to the spin scalar product,
\beq \label{Bsymm}
\Sl \B(x)\, \psi \,|\, \phi \Sr_x = \Sl \psi \,|\, \B(x)\, \phi \Sr_x  \qquad {\mbox{for all~$x \in \scrM$ and~$\psi, \phi \in S_x$}}.
\eeq
Then the Dirac equation becomes
\beq \label{Dout}
(\Dir -m)\, \psi_m = 0 \qquad \text{with} \qquad \Dir := i \Pdd + \B \:.
\eeq
Since the Dirac equation is linear and hyperbolic
(meaning that it can be rewritten as a symmetric hyperbolic system), 
its Cauchy problem for smooth initial data is well-posed,
giving rise to global smooth solutions. Moreover, due to finite propagation speed, starting with
compactly supported initial data, we obtain solutions which are spatially compact at any time.
We denote the space of such smooth wave functions with spatially compact support by~$\Cisc(\scrM, S\scrM)$.
Using the symmetry assumption~\eqref{Bsymm}, for any solutions~$\psi_m, \phi_m \in \Cisc(\scrM,S\scrM)$ of the
Dirac equation the vector field~$\Sl \psi_m | \gamma^j \phi_m \Sr$ is divergence-free;
this is referred to as {\em{current conservation}}. Applying Gauss' divergence theorem,
this implies that the spatial integral
\beq \label{print}
(\psi_m | \phi_m)_m \big|_t := 2 \pi \int_{\R^3} \Sl \psi_m | \gamma^0 \phi_m \Sr|_{(t, \vec{x})}\: d^3x
\eeq
is independent of the choice of the space-like hypersurface labelled by the time parameter~$t$. This integral defines a scalar product on the solution space
corresponding to the mass~$m$. Forming the completion, we obtain a Hilbert space, which we
denote by~$(\H_m, (.|.)_m)$. The norm on~$\H_m$ is~$\| . \|_m$.

The {\em{retarded}} and {\em{advanced Green's operators}}~$\tilde{s}_m^\wedge$ and~$\tilde{s}_m^\vee$ are
mappings (for details see for example~\cite{baer+ginoux})
\[ \tilde{s}_m^\wedge, \tilde{s}_m^\vee \::\: C^\infty_0(\scrM, S\scrM) \rightarrow \Cisc(\scrM, S\scrM)\:. \]
Their difference is the so-called causal fundamental solution~$\tilde{k}_m$,
\beq \label{kmdef}
\tilde{k}_m := \frac{1}{2 \pi i} \left( \tilde{s}_m^\vee - \tilde{s}_m^\wedge \right) \::\: C^\infty_0(\scrM, S\scrM) \rightarrow \Cisc(\scrM, S\scrM)
\cap \H_m \:.
\eeq
These operators can be represented as integral operators with a distributional kernel, for example,
\[ (\tilde{k}_m \phi)(x) = \int_\scrM \tilde{k}_m(x,y)\, \phi(y)\: d^4y\:. \]
Leaving out the tilde always refers to the special case~$\B \equiv 0$.

The operator~$\tilde{k}_m$ can be used for constructing a solution of the Cauchy problem.
To this end, we always work in the foliation~$\scrN_t = \{(t, \vec{x}) \,|\, \vec{x} \in \R^3\}$ of constant time
Cauchy hypersurfaces in a fixed reference frame~$(t, \vec{x})$.
For clarity, we denote the Hilbert space of square integrable
spinors at time~$t$ with the scalar product~\eqref{print} by~$(\H_t, (.|.)|_t)$.
Moreover, we denote a wave function~$\psi$ at time~$t$ by~$\psi|_t$
(we use this notation both for the restriction of a wave function in space-time and for
a function defined only on the hyperplane~$\scrN_t$).
\begin{Prp} \label{prp21} The solution of the Cauchy problem
\beq \label{cauchyt}
(\Dir - m) \,\psi_m = 0 \:,\qquad \psi_m \big|_{t_0} = \psi_0 \in C^\infty(\scrN_{t_0} \simeq \R^3, S\scrM)
\eeq
has the representation
\[ \psi_m(x) = 2 \pi \int_{\scrN_{t_0}} \tilde{k}_m \big(x,(t_0,\vec{y}) \big)\, \gamma^0\, \psi_0(\vec{y})\: d^3y\:. \]
\end{Prp} \noindent
For the proof see for example~\cite[Section~2]{finite}.

Moreover, the operator~$\tilde{k}_m$ can be regarded as the signature operator of the
inner product~\eqref{stip} when expressed in terms of the scalar product~\eqref{print}.
\begin{Prp} \label{prpdual}
For any~$\psi_m \in \H_m$ and~$\phi \in C^\infty_0(\scrM, S\scrM)$,
\beq \label{dual}
(\psi_m \,|\, \tilde{k}_m \phi)_m = \bra \psi_m | \phi \ket \:.
\eeq
\end{Prp} \noindent
For the proof we refer to~\cite[Proposition~2.2]{dimock3} or~\cite[Section~3.1]{finite}.

The unique solvability of the Cauchy problem allows us to introduce
the group of {\em{time evolution operators}} as follows.
According to Proposition~\ref{prp21}, for given initial data~$\psi_0 \in C^\infty_0(\scrN_{t_0}, S\scrM)$,
the Cauchy problem~\eqref{cauchyt} has a unique solution~$\psi_m \in \Cisc(\scrM, S\scrM) \cap \H_m$.
Evaluating this solution at some other time~$t$ gives a mapping~$\tilde{U}_m^{t,t_0} : \psi_0 \mapsto \psi_m|_t$.
Since the scalar product~\eqref{print} is time independent, the time evolution operator~$\tilde{U}_m^{t,t_0}$
is isometric. Thus by continuity, it extends uniquely to an isometry
\[ \tilde{U}_m^{t, t_0} \::\: \H_{t_0} \rightarrow \H_t\:. \]
Since~$t_0$ can be chosen arbitrarily and the Cauchy problem can be solved forward and backward in time,
this isometry is even a unitary operator. Moreover, these operators are a
representation of the group~$(\R, +)$, meaning that
\[ \tilde{U}_m^{t,t} = \1 \qquad \text{and} \qquad \tilde{U}_m^{t'',t'}\: \tilde{U}_m^{t',t} = \tilde{U}_m^{t'',t}\:. \]
Proposition~\ref{prp21} immediately yields the following representation of~$\tilde{U}^{t',t}_m$
with integral kernel,
\begin{align}
\big( \tilde{U}^{t',t}_m \,\psi|_{t} \big)(\vec{y}) &= \int_{\R^3} \tilde{U}^{t',t}_m(\vec{y}, \vec{x})\: 
\psi|_t(\vec{x})\: d^3x\:, 
\label{Ukern} \\
\tilde{U}^{t',t}_m(\vec{y}, \vec{x}) &= 2 \pi \,\tilde{k}_m \big( (t',\vec{y}), (t, \vec{x}) \big)\, \gamma^0\:.
\label{Ukm}
\end{align}

\subsection{The Mass Oscillation Property} \label{secmops}
We denote the families of smooth wave functions with spatially compact support, which are also compactly supported in~$I$, by~$\Cisco(\scrM \times I, S\scrM)$.
The following construction shows that within this class, there are families~$(\psi_m)_{m \in I}$
such that for every~$m \in I$, the wave function~$\psi_m$ is a solution of the
Dirac equation~\eqref{Dirext}: Let~$\psi^0 \in \Cisco(\scrN_{t_0} \times I, S\scrM)$
be a family of smooth and compactly supported functions on~$\scrN \times I$
(for example of the form~$\psi^0(x,m) = \eta(m)\, \chi(x)$ with~$\eta \in C^\infty_0(I)$
and~$\chi \in C^\infty_0(\scrN, S\scrM)$). Solving for every~$m \in I$ the Cauchy problem
\[ (i \Pdd + \B - m) \psi_m = 0 \:, \qquad \psi_m|_\scrN = \psi_m^0 \]
(where again~$\psi_m^0(x) \equiv \psi^0(x,m)$),
we obtain a family~$(\psi_m)_{m \in I}$ of solutions of the Dirac equation
for a variable mass parameter~$m \in I$ in the desired class~$\Cisco(\scrM \times I, S\scrM)$.

The space of families of Dirac solutions in the class~$\Cisco(\scrM \times I, S\scrM)$
are denoted by~$\H^\infty$. On~$\H^\infty$ we introduce the scalar product
\beq \label{scalfamily}
( \psi | \phi) = \int_I (\psi_m | \phi_m)_m \: dm\:,
\eeq
where~$dm$ is the Lebesgue measure (and~$\psi = (\psi_m)_{m \in I}$
and~$\phi = (\phi_m)_{m \in I}$ are families of Dirac solutions for a variable mass parameter).
Forming the completion yields the Hilbert space $(\H, (.|.))$ with norm~$\| . \|$.
Then~$\H^\infty$ can be regarded as the subspace
\beq \label{Hinfchoice}
\H^\infty = \Cisco(\scrM \times I, S\scrM) \cap \H \:.
\eeq
On~$\H$, we introduce the operator of multiplication by~$m$,
\[ T \::\: \H \rightarrow \H \:,\qquad (T \psi)_m = m \,\psi_m \:. \]
Obviously, this operator preserves the support properties, and thus
\[ T|_{\H^\infty} \::\: \H^\infty \rightarrow \H^\infty \:. \]
Moreover, it is a symmetric operator, and it is bounded because the interval~$I$ is, i.e.
\[ T^* = T \in \Lin(\H) \:. \]
Integration of $\psi_m$ over~$m$ gives another operator
\beq \label{pdef}
\p \::\: \H^\infty \rightarrow \Cisc(\scrM, S\scrM)\:,\qquad
\p \psi = \int_I \psi_m\: dm \:.
\eeq
We point out for clarity that~$\p \psi$ no longer satisfies a Dirac equation.
The following notions were introduced in~\cite{infinite}, and we refer the reader
to this paper for more details.

\begin{Def} \label{defWMass Oscillation Property}
The Dirac operator~$\Dir=i \Pdd + \B$ on Minkowski space~$\scrM$ has the {\bf{weak mass oscillation property}} 
in the interval~$I = (m_L, m_R)$ with domain~$\H^\infty$ if the following conditions hold:
\begin{itemize}
\item[(a)] For every~$\psi, \phi \in \H^\infty$, the
function~$\Sl \p \phi | \p \psi \Sr$ is integrable on~$\scrM$. Moreover, there is
a constant~$c=c(\psi)$ such that
\beq \label{mbound}
|\bra \p \psi | \p \phi \ket| \leq c\, \|\phi\| \qquad \textnormal{for all} \: \phi \in \H^\infty \:. 
\eeq
\item[(b)] For all~$\psi, \phi \in \H^\infty$,
\beq \label{mortho}
\bra \p T \psi | \p \phi \ket = \bra \p \psi | \p T \phi \ket \:.
\eeq
\end{itemize}
\end{Def}

\begin{Def} \label{defSMass Oscillation Property}
The Dirac operator~$\Dir=i \Pdd + \B$ on Minkowski space~$\scrM$ has the {\bf{strong mass oscillation property}} in the
interval~$I=(m_L, m_R)$ with domain~$\H^\infty$ if there is a constant~$c>0$ such that
\beq \label{smop}
|\bra \p \psi | \p \phi \ket| \leq c \int_I \, \|\phi_m\|_m\, \|\psi_m\|_m\: dm
\qquad  \textnormal{for all} \: \psi, \phi \in \H^\infty\:.
\eeq
\end{Def}

\noindent The following theorem is proved in~\cite[Theorem~4.2, Proposition~4.3 and
Theorem~4.7]{infinite}.
\begin{Thm} \label{thmSrep}
Assume that the Dirac operator~$\Dir$ has the strong mass oscillation property in the
interval~$I=(m_L, m_R)$. Then there exists a family of linear operators $(\tilde{\Sig}_m)_{m \in I}$
with~$\tilde{\Sig}_m \in \Lin(\H_m)$ which are uniformly bounded,
\[ \sup_{m \in I} \|\tilde{\Sig}_m\| < \infty\:, \]
such that
\beq \label{Smdef}
\bra \p \psi | \p \phi \ket = \int_I (\psi_m \,|\, \tilde{\Sig}_m \,\phi_m)_m\: dm \qquad
\text{for all~$\psi, \phi \in \H^\infty$}\:.
\eeq
The operator~$\tilde{\Sig}_m$ is uniquely determined for every~$m \in I$ by demanding that
for all~$\psi, \phi \in \H^\infty$, the functions~$( \psi_m | \tilde{\Sig}_m \phi_m)_m$ are continuous in~$m$.
Moreover, the operator~$\tilde{\Sig}_m$ is the same for all choices of~$I$ containing~$m$.
Finally, there is a bi-distribution~${\mathcal{P}} \in \frakD'(\scrM \times \scrM)$ such that the operator~$P$ defined by
\beq \label{Pdeftilde}
P := -\chi_{(-\infty, 0)}(\tilde{\Sig}_m)\, \tilde{k}_m \::\: C^\infty_0(\scrM, S\scrM) \rightarrow \H_m
\eeq
has the representation
\beq \label{disrep}
\bra \phi | P \psi \ket = {\mathcal{P}}(\overline{\phi} \otimes \psi) \qquad
\text{\textnormal{for all}~$\phi, \psi \in C^\infty_0(\scrM, S\scrM)$}
\eeq
(where~$\overline{\phi} = \phi^\dagger \gamma^0$ is the usual adjoint spinor).
\end{Thm} \noindent
The operator~$P$ is referred to as the {\bf{fermionic projector}}.
We also use the standard notation with an integral kernel~$P(x,y)$,
\begin{align*}
\bra \phi | P \psi \ket &= \iint_{\scrM \times \scrM} \Sl \phi(x) \,|\, P(x,y) \,\psi(y) \Sr_x \: d^4x\: d^4y \\
(P \psi)(x) &= \int_{\scrM} P(x,y) \,\psi(y) \: d^4y \:,
\end{align*}
where~$P(.\,,.)$ coincides with the distribution~${\mathcal{P}}$ in~\eqref{disrep}.

\subsection{The Lippmann-Schwinger Equation}
The Dirac dynamics can be rewritten in terms of a symmetric operator~$\tilde{H}$. To this end, we multiply
the Dirac equation~\eqref{Dout} by $\gamma^0$ and bring the $t$-derivative separately on one side of the
equation,
\beq \label{Hamilton}
i \partial_t \psi_m = \tilde{H} \psi_m\:, \qquad \text{where} \qquad
\tilde{H} := -\gamma^0 (i \vec{\gamma} \vec{\nabla} + \B - m)
\eeq
(note that~$\gamma^j \partial_j = \gamma^0 \partial_t + \vec{\gamma} \vec{\nabla}$).
We refer to~\eqref{Hamilton} as the Dirac equation in {\em{Hamiltonian form}}.
The fact that the scalar product~\eqref{print} is time independent
implies that for any two solutions~$\phi_m, \psi_m \in \Cisc(\scrM, S\scrM) \cap \H_m$,
\beq \nonumber
0 = \partial_t (\phi_m \,|\, \psi_m)_m\: = i \bigl((\tilde{H} \phi_m \,|\, \psi_m)_m - (\phi_m \,|\, \tilde{H} \psi_m)_m\bigr)\:,
\eeq
showing that the Hamiltonian is a symmetric operator on ${\mathscr{H}}_m$. The Lippmann-Schwinger equation can be used to compare the dynamics in the Minkowski vacuum with the dynamics in the presence of an external potential. We denote the time evolution operator in the Minkowski vacuum by~$U^{t,t_0}_m$.
\begin{Prp} The Cauchy problem~\eqref{cauchyt} has a solution~$\psi_m$ which satisfies the equation
\beq \label{lse}
\psi_m|_t = U^{t,t_0}_m \psi_0 +i \int_{t_0}^t U^{t, \tau}_m \,\big( \gamma^0 \B\: \psi_m \big) \big|_\tau\: d\tau \:,
\eeq
referred to as the {\bf{Lippmann-Schwinger equation}}.
\end{Prp}
\Proof Obviously, the wave function~$\psi_m|_t$ given by~\eqref{lse} has the correct
initial condition at~$t=t_0$. Thus it remains to show that~$\psi_m|_t$ satisfies the Dirac equation.
To this end, we rewrite the Dirac equation in the Hamiltonian form~\eqref{Hamilton}, and
separate the vacuum Hamiltonian~$H$ from the term involving the external potential,
\beq \label{hamex}
(i \partial_t - H)\, \psi_m = -\gamma^0 \B \,\psi_m \qquad \text{with} \qquad
H = -i \gamma^0 \vec{\gamma} \vec{\nabla} + \gamma^0 m \:.
\eeq
Applying the operator~$i \partial_t - H$ to~\eqref{lse}, and observing that the time evolution
operator maps to solutions of the vacuum Dirac equation, only the derivative of the upper
limit of integration contributes,
\[ (i \partial_t - H)\, \psi_m|_t = - U^{t, \tau}_m \:\left( \gamma^0 \B\: \psi_m \right)\big|_{\tau=t}
= -\gamma^0 \B\: \psi_m|_t \:, \]
so that~\eqref{hamex} is indeed satisfied.
\QED

\section{The Mass Oscillation Property in the Minkowski Vacuum} \label{secminkowski}

Since Minkowski space is ultrastatic, it is known from~\cite[Section~5]{infinite}
that the Dirac operator~$i \Pdd$ satisfies the weak and strong mass
oscillation properties. Moreover, the decomposition of the solution space into the positive and negative spectral
subspaces of the fermionic signature operator reduces to the usual frequency splitting
(see~\cite[Theorem~5.1]{infinite}).
We now reproduce these results giving more explicit proofs.
These explicit results and formulas will be essential for the subsequent treatment of
time-dependent external potentials in Section~\ref{secep}.

Basically, the mass oscillation property in the Minkowski vacuum can be proved easily
using Fourier methods.
Here we shall give two different approaches in detail.
The method of the first proof (Section~\ref{secmweak}) is instructive because it gives an intuitive
understanding of ``mass oscillations''.
However, this method only yields the weak mass oscillation property. The second proof (Section~\ref{secmstrong}) is more abstract but also gives the strong mass oscillation property.

\subsection{Proof of the Weak Mass Oscillation Property using Mass Derivatives} \label{secmweak}
We again consider the foliation~$\scrN_t = \{(t, \vec{x}) \,|\, \vec{x} \in \R^3\}$ of constant time Cauchy hypersurfaces in a fixed reference frame~$(t, \vec{x})$ and a variable mass parameter $m$
in the interval~$I=(m_L, m_R)$ with~$m_L, m_R > 0$. The families of solutions~$\psi = (\psi_m)_{m \in I}$ of the Dirac equations~$(i \Pdd - m) \psi_m = 0$ are contained in the Hilbert space~$(\H, (.|.))$ with the scalar product~\eqref{scalfamily}. Moreover, the subspace~$\H^\infty \subset \H$ is given by~\eqref{Hinfchoice}.

For what follows, it is convenient to work with the Fourier transform in space, i.e.
\[ \hat{\psi}(t, \vec{k}) = \int_{\R^3} \psi(t, \vec{x})\: e^{-i \vec{k} \vec{x}}\: d^3x\:,\qquad
 \psi(t, \vec{x}) = \int_{\R^3} \frac{d^3k}{(2 \pi)^3} \:\hat{\psi}(t, \vec{k})\: e^{i \vec{k} \vec{x}} 
 \:. \]
Then a family of solutions~$\psi \in \H^\infty$ has the representation
\beq \label{Frep}
\hat{\psi}_m(t, \vec{k}) = c_+(\vec{k},m) \: e^{-i \omega(\vec{k}, m)\, t} + c_-(\vec{k}, m)
\: e^{i \omega(\vec{k},m)\, t} \qquad \text{for all~$m \in I$}
\eeq
with suitable spinor-valued coefficients~$c_\pm(\vec{k},m)$ and~$\omega(\vec{k},m):=\sqrt{|\vec{k}|^2+m^2}$.
Integrating over the mass parameter, we obtain a superposition of waves oscillating at different
frequencies. Intuitively speaking, this leads to destructive interference for large~$t$,
giving rise to decay in time. This picture can be made precise using integration by parts in~$m$,
as we now explain. Integrating~\eqref{Frep} over the mass and applying the operator~$\p$, \eqref{pdef},
we obtain
\begin{align*}
\p \hat{\psi}(t, \vec{k}) &= \int_I
\left( c_+\: e^{-i \omega t} + c_-\: e^{i \omega t} \right) dm \\
&= \int_I \frac{i}{t \,\partial_m \omega}
\left( c_+\: \partial_m e^{-i \omega t} - c_-\: \partial_m e^{i \omega t} \right) dm \\
&= -\frac{i}{t}  \int_I \left[ \partial_m \Big( \frac{c_+}{\partial_m \omega} \Big)\: e^{-i \omega t}
-\partial_m \Big( \frac{c_-}{\partial_m \omega} \Big)\: e^{i \omega t} \right] dm \:
\end{align*}
\noindent (we do not get boundary terms because $\psi \in \H^\infty$
has compact support in $m$). With $\partial_m \omega = m/\omega$, we conclude that
\[ \p \hat{\psi}(t, \vec{k}) = -\frac{i}{t}  \int_I \left[ \partial_m \Big( \frac{\omega\, c_+}{m} \Big)\:
e^{-i \omega t} -\partial_m \Big( \frac{\omega\, c_-}{m} \Big)\: e^{i \omega t} \right] dm \:. \]
Since the coefficients~$c_\pm$ depend smoothly on~$m$, the resulting integrand is
bounded uniformly in time, giving a decay at least like~$1/t$, i.e.~$|\p \hat{\psi}(t, \vec{k})| \lesssim 1/t$.
Iterating this procedure, one even can prove decay rates~$\lesssim 1/t^2, 1/t^3, \ldots$
The price one pays is that higher and higher powers in~$\omega$ come up in the integrand,
which means that in order for the spatial Fourier integral to exist, one needs a faster decay of~$c_\pm$ in~$|\vec{k}|$. Expressed in terms of the initial data, this means that every factor~$1/t$ gives rise to an additional spatial derivative acting on the initial data. This motivates the following basic estimate.

\begin{Lemma} \label{lemmappsi}
For any~$\psi \in \H^\infty$, there is a constant~$C=C(m_L)$ such that
\beq \label{1es}
\big\| (\p \psi)|_t \big\|_t \leq \frac{C\,|I|}{1+t^2} \:
\sup_{m \in I} \,\sum_{b=0}^2 \big\| (\partial_m^b \psi_m)|_{t=0} \big\|_{W^{2,2}} \:,
\eeq
where~$\|.\|_t$ is the norm corresponding to the scalar product
\[ ( . | .)|_t := 2 \pi \int_{\R^3} \Sl \,.\, | \gamma^0 \,.\, \Sr_{\vec{x}}\: d^3x \::\:
L^2(\scrN_t, S\scrM) \times L^2(\scrN_t, S\scrM) \rightarrow \C \]
(which is similar to~\eqref{print}, but now applied to wave functions which do not need
to be solutions), and~$\| . \|_{W^{2,2}}$ is the spatial Sobolev norm
\beq \label{sobolev}
\| \phi \|^2_{W^{2,2}} := \sum_{\text{$\alpha$ with~$|\alpha| \leq 2$}}\:
\int_{\R^3} | \nabla^\alpha \phi(\vec{x}) |^2\: d^3x \:,
\eeq
where~$\alpha$ is a multi-index.
\end{Lemma} \noindent
The absolute value in~\eqref{sobolev} is the norm~$|\,.\,| := \sqrt{\Sl . | \gamma^0 . \Sr}$
on the spinors. If we again identify all spinor spaces in the Dirac representation with~$\C^4$,
this simply is the standard Euclidean norm on~$\C^4$.

The proof of this lemma will be given later in this section.
Before, we infer the weak mass oscillation property.
\begin{Corollary} \label{corwMOP}
The vacuum Dirac operator~$i \Pdd$ in Minkowski space has the weak mass oscillation property
with domain~\eqref{Hinfchoice}.
\end{Corollary}
\Proof
For every~$\psi, \phi \in \H^\infty$, the Schwarz inequality gives
\beq \label{schwarz}
|\bra \p \psi | \p \phi \ket| = \frac{1}{2 \pi}
\left| \int_{-\infty}^\infty \big( (\p \psi)|_t \,\big|\, \gamma^0 \,(\p \phi)|_t \big)_t \:dt \right| 
\leq \int_{-\infty}^\infty \big\| (\p \psi)|_t \big\|_t\: \big\| (\p \phi)|_t \big\|_t\: dt \:.
\eeq
Applying Lemma~\ref{lemmappsi} together with the estimate
\begin{align*}
\big\| (\p \phi)|_t \big\|^2_t &= \iint_{I \times I} \big( \phi_m|_t \,\big|\, \phi_{m'}|_t \big)_t \:dm \,dm' \\
&\leq \frac{1}{2} \iint_{I \times I} \Big( \|\phi_m\|_m^2 +  \|\phi_{m'}\|_{m'}^2 \Big)\: dm \,dm' = |I| \,\|\phi\|^2 \:,
\end{align*}
we obtain inequality~\eqref{mbound} with
\beq \label{ces}
c = C\,|I|^\frac{3}{2}\:
\sup_{m \in I} \,\sum_{b=0}^2 \|\partial_m^b (\psi_m)|_{t=0} \|_{W^{2,2}}
\int_{-\infty}^\infty \frac{1}{1+t^2} \: dt < \infty \:.
\eeq
 
The identity~\eqref{mortho} follows by integrating the Dirac operator by parts,
\beq \label{intpart}
\begin{split}
\bra \p T \psi | \p \phi \ket &= \bra \p \Dir \psi | \p \phi \ket =
\bra \Dir \p \psi | \p \phi \ket = \int_\scrM \Sl \Dir \p \psi | \p \phi \Sr_x\: d^4x \\
&\overset{(\star)}{=} \int_\scrM \Sl \p \psi | \Dir \p \phi \Sr_x\: d^4x = \bra \p \psi | \Dir \p \phi \ket
= \bra \p \psi | \p T \phi \ket \:.
\end{split}
\eeq
In~$(\star)$, we used that the Dirac operator is formally self-adjoint with respect to the inner product~$\bra .|. \ket$.
Moreover, we do not get boundary terms because of the time decay in Lemma~\ref{lemmappsi}.
\QED

The remainder of this section is devoted to the proof of Lemma~\ref{lemmappsi}.
Specializing the result of Proposition~\ref{prp21} to the Minkowski vacuum,
we can express the solution~$\psi_m$ of the Cauchy problem
in terms of the causal fundamental solution $k_m$. In order to bring~$k_m$ into a more
explicit form, we use~\eqref{kmdef} together with formulas for the advanced and retarded Green's functions.
Indeed, these Green's functions are the multiplication operators in momentum space
\[ s^{\lor}_{m}(k) = \lim_{\varepsilon \searrow 0}
            \frac{\slashed{k} + m}{k^{2}-m^{2}-i \varepsilon k^{0}}
            \qquad {\mbox{and}} \qquad
           s^{\land}_{m}(k) = \lim_{\varepsilon \searrow 0}
            \frac{\slashed{k} + m}{k^{2}-m^{2}+i \varepsilon k^{0}} \]
(with the limit~$\varepsilon \searrow 0$ taken in the distributional sense, and where the vector~$k$ is 
the four-momentum). We thus obtain in momentum space
\begin{align*}
k_{m}(p) &=\frac{1}{2 \pi i} \:(\slashed{p} + m) \; \lim_{\varepsilon
\searrow 0} \left[ \frac{1}{p^{2}-m^{2}-i\varepsilon p^{0}} \:-\:
\frac{1}{p^{2}-m^{2}+i\varepsilon p^{0}} \right] \\
&= \frac{1}{2 \pi i} \:(\slashed{p} + m) \; \lim_{\varepsilon
\searrow 0} \left[ \frac{1}{p^{2}-m^{2}-i\varepsilon} \:-\:
\frac{1}{p^{2}-m^{2}+i\varepsilon} \right] \epsilon(p^{0})
\end{align*}
(where~$\epsilon$ denotes the step function, and for notational clarity we denoted the
momentum variables by~$p$).
Employing the distributional equation
\[ \lim_{\varepsilon \searrow 0} \left( \frac{1}{x - i \varepsilon} - \frac{1}{x + i \varepsilon} \right)
= 2 \pi i \: \delta(x) \:, \]
we obtain the simple formula
\beq \label{kmp}
k_m(p) = (\slashed{p} + m) \: \delta(p^{2}-m^{2}) \: \epsilon(p^{0}) \:.
\eeq
It is convenient to transform spatial coordinates of the time evolution operator to momentum space. 
First, in the Minkowski vacuum, the time evolution operator can be represented as in~\eqref{Ukern} with
an integral kernel~$U^{t,t'}(\vec{y}, \vec{x})$ which depends only on the difference vector~$\vec{y}-\vec{x}$.
We set
\[ U^{t,t'}(\vec{k}) := \int_{\R^3} U^{t,t'}(\vec{y}, 0) \:e^{-i \vec{k} \vec{y}}\: d^3y \:. \]
Combining~\eqref{Ukm} with~\eqref{kmp} yields
\[ U^{t,t'}(\vec{k}) = \int_{-\infty}^\infty (\slashed{k} + m)\:\gamma^0 \: \delta(k^{2}-m^{2}) \big|_{k=(\omega, \vec{k})}\;
\epsilon(\omega)\: e^{-i \omega (t-t')}\: d\omega \:. \]
Carrying out the $\omega$-integral, we get
\beq \label{Urep}
U^{t,t'}(\vec{k}) = \sum_{\pm} \Pi_\pm(\vec{k}) \:e^{\mp i \omega (t-t')} \:,
\eeq
where we set
\begin{gather}
\Pi_\pm(\vec{k}) := \pm \frac{1}{2 \omega(\vec{k})} \; (\slashed{k}_\pm + m) \,\gamma^0 \label{Pidef2} \\
\text{with} \qquad \omega(\vec{k}) = \sqrt{|\vec{k}|^2 + m^2}
\qquad \text{and} \qquad k_\pm = (\pm \omega(\vec{k}), \vec{k} )\:. \nonumber
\end{gather}
Moreover, applying Plancherel's theorem, the scalar product~\eqref{print} can be written in momentum space as
\[ (\psi_m \,|\, \phi_m)_m = (2 \pi)^{-2} \int_{\R^3} \Sl \hat{\psi}_m(t, \vec{k}) \,|\,
\gamma^0 \,\hat{\phi}_m(t, \vec{k}) \Sr\: d^3k \:. \]
The unitarity of the time evolution operator in position space implies that the matrix~$U^{t,t'}(\vec{k})$ is
unitary (with respect to the scalar product~$\la .\,,. \ra_{\C^2} \equiv \Sl \,.\,| \gamma^0 \,.\, \Sr$), meaning that its eigenvalues are
on the unit circle and the corresponding eigenspaces are orthogonal.
It follows that the operators~$\Pi_\pm(\vec{k})$ in~\eqref{Urep} are the orthogonal projection operators to the
eigenspaces corresponding to the eigenvalues~$e^{\mp i \omega (t-t')}$, i.e.
\[ \gamma^0 \Pi_s^* \gamma^0 = \Pi_s \quad \text{and} \quad \Pi_s(\vec{k})\: \Pi_{s'}(\vec{k}) = \delta_{s, s'}\: \Pi_s(\vec{k}) \qquad \text{for~$s,s' \in \{+,-\}$}\:. \]
(these relations can also be verified by straightforward computations using~\eqref{Pidef2}).

The next two lemmas involve derivatives with respect to the mass parameter~$m$.
For clarity, we again denote the $m$-dependence of the operators by the subscript~$m$.
\begin{Lemma} \label{lemmamint}
The time evolution operator in the vacuum satisfies the relation
\begin{align}
(t-t')\, U_m^{t,t'}(\vec{k}) &= \frac{\partial}{\partial m} V_m^{t,t'}(\vec{k}) + W_m^{t,t'}(\vec{k}) \:, \label{tes}
\intertext{where}
V_m^{t,t'}(\vec{k}) &= \sum_\pm \frac{i}{2m}\: (\slashed{k}_\pm + m) \gamma^0  \:e^{\mp i \omega (t-t')}
\label{Vmdef} \\
W_m^{t,t'}(\vec{k}) &= \sum_\pm \frac{i}{2} \Big( \frac{\slashed{k}_\pm \gamma^0}{m^2}
\mp \frac{1}{\omega} \Big)  \:e^{\mp i \omega (t-t')} \:. \label{Wmdef}
\end{align}
The operators~$V_m^{t,t'}$ and~$W_m^{t,t'}$ are estimated uniformly by
\beq \label{VWes}
\|V_m^{t,t'}(\vec{k})\| + \|W_m^{t,t'}(\vec{k})\| \leq C \,\bigg( 1 + \frac{|\vec{k}|}{m} \bigg) \:,
\eeq
where the constant~$C$ is independent of~$m$, $\vec{k}$, $t$ and~$t'$
(and~$\| \,.\, \|$ is any norm on the $2\times 2$-matrices).
\end{Lemma}
\Proof First, we generate the factor~$t-t'$ by differentiating the exponential in~\eqref{Urep} with
respect to~$\omega$,
\begin{align*}
(t-t')\, U_m^{t,t'}(\vec{k}) &= \sum_\pm
\Pi_\pm(\vec{k}) \Big( \pm i \frac{\partial}{\partial \omega}\:e^{\mp i \omega (t-t')} \Big).
\end{align*}
Next, we want to rewrite the $\omega$-derivative as a derivative with respect to~$m$.
Taking the total differential of the dispersion relation~$\omega^2 - |\vec{k}|^2 = m^2$ for fixed $\vec{k}$, one finds that
\beq \label{omegam}
\frac{\partial}{\partial \omega} = \frac{\omega}{m}\: \frac{\partial}{\partial m} \:.
\eeq
Hence
\begin{align*}
(t-t')\,U_m^{t,t'} &= \sum_\pm
\Pi_\pm \Big( \pm i \:\frac{\omega}{m} \frac{\partial}{\partial m}\:e^{\mp i \omega (t-t')} \Big) \\
&= \frac{\partial}{\partial m} \sum_\pm \left( \pm i \:\frac{\omega}{m} \:
\Pi_\pm \:e^{\mp i \omega (t-t')} \right)
- \sum_\pm \left( \frac{\partial}{\partial m} \left[ \pm i \:\frac{\omega}{m} \:
\Pi_\pm \right] \right) e^{\mp i \omega (t-t')}\:.
\end{align*}
Computing the operators in the round brackets using~\eqref{Pidef2} gives the identities~\eqref{Vmdef}
and~\eqref{Wmdef}. Estimating these formulas, one obtains bounds which are at most linear
in~$|\vec{k}|$, proving~\eqref{VWes}.
\QED

This method can be iterated to generate more factors of~$t-t'$.
In the next lemma, we prove at least quadratic decay in time.
For later use, it is preferable to formulate the result in position space.
\begin{Lemma} \label{lemmaint2}
The time evolution operator in the vacuum has the representation
\beq \label{Udecay}
U_m^{t,t'} = \frac{1}{(t-t')^2} \left( \frac{\partial^2}{\partial m^2} A_m^{t,t'}+ 
\frac{\partial}{\partial m} B_m^{t,t'} + C_m^{t,t'} \right)
\eeq
with operators
\[ A_m^{t,t'}, B_m^{t,t'}, C_m^{t,t'} \::\: W^{2,2}(\scrN_{t'}, S\scrM) \rightarrow L^2(\scrN_t, S\scrM)\:, \]
which are bounded uniformly in time by
\beq \label{ABCsob}
\|A_m^{t,t'}(\phi)\|_t + \|B_m^{t,t'}(\phi)\|_t + \|C_m^{t,t'}(\phi)\|_t
\leq c\, \|\phi\|_{W^{2,2}} \:,
\eeq
where~$c$ is a constant which depends only on~$m$.
\end{Lemma}
\Proof A straightforward computation using exactly the same methods as
in Lemma~\ref{lemmamint} yields the representation
\beq \label{t2es}
(t-t')^2\, U_m^{t,t'}(\vec{k}) = \frac{\partial^2}{\partial m^2} A_m^{t,t'}(\vec{k}) + 
\frac{\partial}{\partial m} B_m^{t,t'}(\vec{k}) + C_m^{t,t'}(\vec{k}) \:,
\eeq
where the operators~$A_m^{t,t'}$, $B_m^{t,t'}$ and~$C_m^{t,t'}$ are bounded by
\beq \label{ABCes}
\|A_m^{t,t'}(\vec{k})\| + \|B_m^{t,t'}(\vec{k})\| + \|C_m^{t,t'}(\vec{k})\| 
\leq \frac{C}{m} \:\bigg( 1 + \frac{|\vec{k}|}{m} + \frac{|\vec{k}|^2}{m^2} \bigg) \:,
\eeq
with a numerical constant~$C>0$. We remark that, compared to~\eqref{tes}, the right of~\eqref{ABCes}
involves an additional $1/m$. This prefactor
is necessary for dimensional reasons, because the additional factor~$t-t'$ in~\eqref{t2es}
(compared to~\eqref{tes}) brings in an additional dimension of length
(and in natural units, the factor~$1/m$ also has the dimension of length).
The additional summand $|\vec{k}|^2/m^2$ in~\eqref{ABCes} can be understood from the fact that applying~\eqref{omegam} generates a factor of~$\omega/m$ which for large~$|\vec{k}|$
scales like~$|\vec{k}|/m$.

Translating this result to position space and keeping in mind that the vector~$\vec{k}$ corresponds to the derivative~$-i\vec{\nabla}$, we obtain the result.
\QED

\Proof[Proof of Lemma~\ref{lemmappsi}]
First of all, the Schwarz inequality gives
\[ \big\| (\p \psi)|_t \big\|_t \leq \int_I \|\psi_m\|_m\, dm \leq \sqrt{|I|} \;\|\psi\| \:. \]
Thus it remains to show the decay for large~$t$, i.e.
\beq \label{es1}
\big\| (\p \psi)|_t \big\|_t \leq \frac{C\,|I|}{t^2} \:
\sup_{m \in I} \,\sum_{b=0}^2 \|\partial_m^b (\psi_m)|_{t=0} \|_{W^{2,2}} \:.
\eeq

We apply Lemma~\ref{lemmaint2} and integrate by parts in~$m$ to obtain
\begin{align*}
(\p \psi)|_t &= \int_I U_m^{t,0} \:\psi_m|_{t=0}\: dm = \frac{1}{t^2} \int_I \big( \partial^2_m A_m^{t,0}
+ \partial_m B_m^{t,0} + C_m^{t,0} \big)\:\psi_m|_{t=0}\: dm \\
&= \frac{1}{t^2} \int_I \Big( A_m^{t,0}\:  (\partial^2_m \psi_m|_{t=0})
- B_m^{t,0}\:  (\partial_m \psi_m|_{t=0}) + C_m^{t,0}\: \psi_m|_{t=0} \Big)\: dm \:.
\end{align*}
Taking the norm and using~\eqref{ABCsob} gives~\eqref{es1}.
\QED

We finally note that the previous estimates are not optimal for two reasons. First, the pointwise
quadratic decay in~\eqref{1es} is more than what is needed for the convergence of the integral in~\eqref{ces}.
Second and more importantly, the Schwarz inequality~\eqref{schwarz} does not catch the optimal
scaling behavior in~$\vec{k}$. This is the reason why the constant in~\eqref{mbound}
involves derivatives of~$\psi_m$ (cf.~\eqref{ces}), making it impossible to prove inequality~\eqref{smop}
which arises in the strong mass oscillation property.
In order to improve the estimates, one needs to use Fourier methods both in space and time,
as will be explained in the next section.

\subsection{Proof of the Mass Oscillation Property using a Plancherel Method} \label{secmstrong}

\begin{Thm} \label{thmsMOP}
The vacuum Dirac operator in Minkowski space has the strong mass oscillation property
with domain~\eqref{Hinfchoice}.
\end{Thm} \noindent
Our proof relies on a Plancherel argument in space-time. It also provides an
alternative method for establishing the weak mass oscillation property.
\Proof[Proof of Theorem~\ref{thmsMOP}]
Let~$\psi=(\psi_m)_{m \in I} \in \H^\infty$ be a family of solutions of the Dirac equation
for a varying mass parameter in the Minkowski vacuum.
Using Proposition~\ref{prp21}, one can express~$\psi_m$ in terms of its values at time~$t=0$ by
\[ \psi_m(x) = 2 \pi \int_{\R^3} k_m(x, (0, \vec{y})) \,\gamma^0\, \psi_m|_{t=0}(\vec{y})\: d^3y \:. \]
We now take the Fourier transform, denoting the four-momentum by~$k$.
Using~\eqref{kmp}, we obtain
\begin{align*}
\psi_m(k) &= 2 \pi k_m(k)\, \gamma^0 \hat{\psi}_m^0(\vec{k}) \\
&= 2 \pi \: \delta(k^2-m^2)\: \epsilon(k^0)\: (\slashed{k}+m) \,\gamma^0 \hat{\psi}_m^0(\vec{k}) \:,
\end{align*}
where~$\hat{\psi}_m^0(\vec{k})$ denotes the spatial Fourier transform of~$\psi_m|_{t=0}$.
Obviously, this is a distribution supported on the mass shell. In particular, it is
not square integrable over~$\R^4$.

Integrating over~$m$, we obtain the following function
\beq \label{ppsi}
(\p \psi)(k) = 2 \pi \: \chi_I(m)\: \frac{1}{2m}\: \epsilon(k^0)\: (\slashed{k}+m) \,\gamma^0 
\hat{\psi}_m^0(\vec{k})
\Big|_{m=\sqrt{k^2}} \:,
\eeq
where~$m$ now is a function of the momentum variables.
Since the function~$\psi_m|_{t=0}$ is compactly supported and smooth in the spatial variables,
its Fourier transform~$\hat{\psi}_m^0(\vec{k})$
has rapid decay. This shows that the function~\eqref{ppsi} is indeed square integrable.
Using Plancherel, we see that condition~(a) in Definition~\ref{defWMass Oscillation Property} is satisfied.
Moreover, the operator~$T$ is simply the operator of multiplication by~$\sqrt{k^2}$,
so that condition~(b) obviously holds. This again shows the weak mass oscillation property.

In order to prove the strong mass oscillation property, we
need to compute the inner product~$\bra \p \psi | \p \phi \ket$.
To this end, we first write this inner product in momentum space as
\begin{align*}
\bra \p \psi | \p \phi \ket &= \int \frac{d^4k}{(2 \pi)^4}
4 \pi^2 \: \chi_I(m)\: \frac{1}{4m^2}\: \Sl (\slashed{k}+m) \,\gamma^0 \hat{\psi}_m^0(\vec{k}) \,|\,
(\slashed{k}+m) \,\gamma^0 \hat{\phi}_m^0(\vec{k}) \Sr \Big|_{m=\sqrt{k^2}} \\
&= \int \frac{d^4k}{4 \pi^2}\:
\chi_I(m)\: \frac{1}{2m}\: \Sl \gamma^0 \hat{\psi}_m^0(\vec{k}) \,|\,
(\slashed{k}+m) \,\gamma^0 \hat{\phi}_m^0(\vec{k}) \Sr \Big|_{m=\sqrt{k^2}} \:.
\end{align*}
Reparametrizing the $k^0$-integral as an integral over~$m$, we obtain
\beq \label{goback}
\bra \p \psi | \p \phi \ket
= \frac{1}{4 \pi^2} \int_I dm \int_{\R^3} \frac{d^3k}{2\, |k^0|}\: \Sl \gamma^0 \hat{\psi}_m^0(\vec{k}) \,|\,
(\slashed{k}+m) \,\gamma^0 \hat{\phi}_m^0(\vec{k}) \Sr \big|_{k^0 = \pm \sqrt{|\vec{k}|^2+m^2}} \:.
\eeq
Estimating the inner product with the Schwarz inequality and applying Plancherel's theorem, one finds
\[ |\bra \p \psi | \p \phi \ket| \leq \frac{1}{4 \pi^2} \int_I dm
\int_{\R^3} \|\hat{\psi}_m^0(\vec{k})\|\, \| \hat{\phi}_m^0(\vec{k})\| \: d^3k
\leq 2 \pi \int_I \|\psi_m\|_m\: \|\phi_m\|_m \: dm \:. \]
Thus the inequality~\eqref{smop} holds.
\QED

\section{The Mass Oscillation Property in Minkowski Space with External Potential} \label{secep}
\subsection{Proof of the Weak Mass Oscillation Property}
In this section, we prove the following theorem.
\begin{Thm} \label{thm1}
Assume that the time-dependent external potential~$\B$ is smooth and
decays faster than quadratically for large times in the sense that~\eqref{Bdecay} 
holds for suitable constants~$c, \varepsilon>0$.
Then the Dirac operator~$\Dir = i \Pdd + \B$ has the weak mass oscillation property.
\end{Thm}
\noindent We expect that this theorem could be improved by weakening the decay assumptions
on the potential.
However, this would require refinements of our methods which would go beyond the scope of this paper.
Also, using that Dirac solutions dissipate, the pointwise decay in time could probably be replaced
or partially compensated by suitable spatial decay assumptions.
Moreover, one could probably refine the
result of the above theorem by working with other norms (like weighted~$C^k$- or Sobolev norms).

The main step is the following basic estimate, which is the analog of Lemma~\ref{lemmappsi}
in the presence of an external potential.
\begin{Prp} \label{prpbasic}
Under the decay assumptions~\eqref{Bdecay} on the external potential~$\B$,
there are constants~$c, \varepsilon>0$ such that for every family~$\psi \in \H^\infty$
of solutions of the Dirac equation~\eqref{Dout} with varying mass,
\beq \label{ppsies}
\big\| ( \p \psi \big)|_t \big\|_t \leq \frac{c}{1 + |t|^{1+\varepsilon}} \: \sup_{m \in I}\:
\sum_{b=0}^2  \big\| (\partial_m^b \psi_m)|_{t=0} \big\|_{W^{2,2}} \:.
\eeq
\end{Prp} \noindent

We first show that this proposition implies the weak mass oscillation property.
\Proof[Proof of Theorem~\ref{thm1} under the assumption that Proposition~\ref{prpbasic} holds]
In order to derive the inequality~\eqref{mbound}, we begin with the estimate
\begin{align*}
|\bra \p \psi | \p \phi \ket| &\leq  \frac{1}{2 \pi} \int_{-\infty}^\infty \Big| \big( \p \psi|_t \,\big|\, \p \phi|_t \big) \big|_t \Big|\, dt
\leq \sup_{t \in \R} \big\|\p \phi|_t \big\|_t \int_{-\infty}^\infty \big\| \p \psi|_t \big\|_t \, dt \:.
\end{align*}
The last integral is finite by Proposition~\ref{prpbasic}. The supremum can be bounded
by the Hilbert space norm using the H\"older inequality,
\[ \|\p \phi|_t\|_t = \bigg\| \int_I \phi_m|_t\: dm \bigg\|_t
\leq \int_I \big\| \phi_m|_t \big\|_t\: dm \leq \sqrt{|I|} \left( \int_I \|\phi_m|_t\|_t^2 \: dm \right)^\frac{1}{2}
= \sqrt{|I|} \:\|\phi\| \:, \]
giving~\eqref{mbound}.

Using~\eqref{Bsymm}, the Dirac operator~$i \Pdd+ \B$ is formally self-adjoint with respect
to the inner product~$\bra .| .\ket$. Therefore, the
identity~\eqref{mortho} can be obtained just as in~\eqref{intpart}
by integrating the Dirac operator in space-time by parts,
noting that we do not get boundary terms in view of the time decay in
Proposition~\ref{prpbasic}.
\QED

The remainder of this section is devoted to the proof of Proposition~\ref{prpbasic}.
One ingredient is the Lippmann-Schwinger equation~\eqref{lse},
\beq \label{lse2}
\psi_m|_t = U_m^{t,0} \:\psi_m|_{t=0} +i \int_0^t U_m^{t, \tau} \big( \gamma^0 \B\: \psi_m \big) \big|_\tau\, d\tau \:.
\eeq
Since the first summand of this equation is controlled by Lemma~\ref{lemmappsi},
it remains to estimate the second summand.
Again using~\eqref{Udecay} and integrating by parts with respect to the mass, we obtain
\[ \int_I U_m^{t, \tau} \big( \gamma^0 \B\: \psi_m \big) \big|_\tau\: dm =
\frac{1}{(t-\tau)^2} \int_I \big( A_m^{t,\tau}\:\partial^2_m  -B_m^{t,\tau}  \partial_m + C_m^{t,\tau} \big)
\big( \gamma^0 \B\: \psi_m \big) \big|_\tau \:dm \]
and thus
\begin{align*}
\bigg\| \int_I U_m^{t, \tau} \big( \gamma^0 \B\: \psi_m \big) \big|_\tau\: dm \bigg\|_t
&\leq \frac{c\,|I|}{(t-\tau)^2} \:\sup_{m \in I} \:\sum_{b=0}^2 \big\| \B(\tau)\: (\partial_m^b \psi_m)|_\tau \big\|_{W^{2,2}}
\nonumber \\
&\leq \frac{c\,|I|}{(t-\tau)^2} \:|\B(\tau)|_{C^2}\: \sup_{m \in I}\,
\sum_{b=0}^2 \big\| \partial_m^b \psi_m|_\tau \big\|_{W^{2,2}} \:.
\end{align*}
We now bound~$\B(\tau)$ with the help of~\eqref{Bdecay} and estimate the
Sobolev norm~$\big\| \partial_m^b \psi_m|_\tau \big\|_{W^{2,2}}$
at time~$\tau$ by means of Lemma~\ref{lemmaB} proved in Appendix~\ref{appB}.
This gives rise to the inequality
\[ \bigg\| \int_I U_m^{t, \tau} \big( \gamma^0 \B\: \psi_m \big) \big|_\tau\: dm \bigg\|_t
\leq \frac{c^2 \,C\,|I|}{(t-\tau)^2} \:\frac{1+|\tau|^2}{1+|\tau|^{2+\varepsilon}} \:
\sup_{m \in I}\,\sum_{b=0}^2 \big\| \partial^b_m \psi_m|_{t=0} \big\|_{W^{2, 2}} \:, \]
which yields the desired decay provided that~$\tau$ and~$t$ are not close to each other.
More precisely, we shall apply this inequality in the case~$|\tau| \leq |t|/2$. Then the estimate simplifies to
\begin{align}
\bigg\| \int_I U_m^{t, \tau} \big( &\gamma^0 \B\: \psi_m \big) \big|_\tau\: dm \bigg\|_t \notag \\
& \;\;\,\leq \frac{\tilde{C}}{t^{2}\, (1+|\tau|^\varepsilon)}
\sup_{m \in I}\,\sum_{b=0}^2 \big\| \partial^b_m \psi_m|_{t=0} \big\|_{W^{2, 2}} \qquad
\text{if~$|\tau| \leq |t|/2$} \label{es2}
\end{align}
with a new constant~$\tilde{C}>0$.
In the remaining case~$|\tau| > |t|/2$, we use the unitarity of~$U_m^{t, \tau}$ to obtain
\[ \bigg\| \int_I U_m^{t, \tau} \left( \gamma^0 \B\: \psi_m \right) |_\tau\: dm \bigg\|_t
\leq |I| \: |\B(\tau)|_{C^0}\: \sup_{m \in I} \|\psi_m\|\:. \]
Applying~\eqref{Bdecay} together with the inequality~$|\tau| > |t|/2$, this gives
\beq \label{es3}
\bigg\| \int_I U_m^{t, \tau} \left( \gamma^0 \B\: \psi_m \right) |_\tau\: dm \bigg\|_t
\leq \frac{\tilde{C}}{t^{2+\varepsilon}}\: \sup_{m \in I} \|\psi_m\| \qquad\;\; \text{if~$|\tau| > |t|/2$}\:.
\eeq
This again decays for large~$t$ because~$\tau$ is close to~$t$ and~$|\B(\tau)|_{C^0}$
decays for large~$\tau$.

Comparing~\eqref{es2} and~\eqref{es3}, we find that the inequality in~\eqref{es2}
even holds for all~$\tau$. Thus integrating this inequality over~$\tau \in [0, t]$, we obtain
the following estimate for the second summand in~\eqref{lse2},
\[ \left\| \int_I dm \int_0^t U_m^{t, \tau} \left( \gamma^0 \B\: \psi_m \right) |_\tau\: d\tau \right\|_t
\leq \frac{C'}{t^{1+\varepsilon}}
\sup_{m \in I}\,\sum_{b=0}^2 \big\| \partial^b_m \psi|_{t=0} \big\|_{W^{2, 2}} \]
(where~$C'>0$ is a new constant).
Combining this inequality with the estimate~\eqref{1es} of the first summand in~\eqref{lse2},
we obtain the desired inequality~\eqref{ppsies}. This concludes the proof of Proposition~\ref{prpbasic}.

\subsection{Proof of the Strong Mass Oscillation Property} \label{secproofmass}
In this section, we prove the following result.

\begin{Thm} \label{thmC2} Assume that the weak mass oscillation property holds and
that the external potential~$\B$ satisfies the condition
\beq \label{L1B}
\int_{-\infty}^\infty |\B(\tau)|_{C^0} \:d\tau < \infty \:.
\eeq
Then the Dirac operator~$\Dir = i \Pdd + \B$ has the strong mass oscillation property.
\end{Thm} \noindent
Combining this theorem with Theorem~\ref{thm1}, one immediately obtains Theorem~\ref{thmmop}.

For the proof we shall derive an explicit formula for the fermionic signature operator
(Proposition~\ref{prpSrep}). This formula is obtained by 
comparing the dynamics in the presence of the external potential with that in the Minkowski vacuum
using the Lippmann-Schwinger equation, and by employing distributional relations for products of
fundamental solutions and Green's functions (Lemma~\ref{lemmadistrel}).

We first return to the formula~\eqref{goback} in the Minkowski vacuum.
Applying Plancherel's theorem and using~\eqref{print}, we conclude that
\beq \label{pmk}
\bra \p \psi | \p \phi \ket = \int_I ( \psi_m^0 \,|\, \Sig_m(\vec{k})\, \phi_m^0) \: dm \:,
\eeq
where
\beq \label{S0def}
\Sig_m(\vec{k}) := \sum_{k^0 = \pm \omega(\vec{k})} \frac{\slashed{k}+m}{2\, \omega(\vec{k})}\: \gamma^0 
= \frac{\vec{k} \vec{\gamma}+m}{\omega(\vec{k})}\: \gamma^0 \:.
\eeq
Comparing~\eqref{pmk} with~\eqref{Smdef}, one sees that the matrix~$\Sig_m(\vec{k})$
is indeed the fermionic signature operator, considered as a multiplication operator in momentum space.
By direct computation, one verifies that the matrix~$\Sig_m(\vec{k})$ has eigenvalues~$\pm 1$.

In order to compare the dynamics in the presence of the external potential with that in the Minkowski vacuum,
we work with the Hamiltonian formulation.
We decompose the Dirac Hamiltonian~\eqref{Hamilton} into the Hamiltonian in the Minkowski
vacuum~\eqref{hamex} plus a potential,
\[ \tilde{H} = H + \V \qquad \text{with} \qquad \V := -\gamma^0 \B \:. \]

\begin{Prp}  \label{prpSrep} Assume that the potential~$\B$ satisfies the condition~\eqref{L1B}. 
Then for every~$\psi, \phi \in \H^\infty$,
\beq \label{Sdef2}
\bra \p \psi | \p \phi \ket = \int_I ( \psi_m \,|\, \tilde{\Sig}_m\, \phi_m)_m \: dm\:,
\eeq
where~$\tilde{\Sig}_m : \H_m \rightarrow \H_m$ are bounded linear operators which act
on the wave functions at time~$t_0$ by
\begin{align}
\tilde{\Sig}_m &= \Sig_m -\frac{i}{2} \int_{-\infty}^\infty \epsilon(t-t_0)\,
\big[ \Sig_m \,U^{t_0, t}_m\, \V(t)\, \tilde{U}^{t,t_0}_m
-\tilde{U}^{t_0, t}_m\, \V(t)\, \Sig_m \,U^{t,t_0}_m \big] \:dt \label{Sm1} \\
&\quad +\frac{1}{2} \left( \int_{t_0}^\infty \!\!\!\int_{t_0}^\infty + \int_{-\infty}^{t_0} \int_{-\infty}^{t_0} \right)
\tilde{U}^{t_0, t}_m\, \V(t) \,\Sig_m\,U^{t, t'}_m\, \V(t')\, \tilde{U}^{t',t_0}_m\:dt\, dt' \label{Sm2}
\end{align}
(and $\Sig_m$ is again the fermionic signature operator of the vacuum~\eqref{S0def}).
\end{Prp}

\noindent Before entering the proof of this proposition, it is instructive to verify
that the above formula for~$\tilde{\Sig}_m$ does not depend on the choice of~$t_0$.
\begin{Remark} {\bf{(Independence of~$\tilde{\Sig}_m$ on~$t_0$)}} {\em{
Our strategy is to differentiate the above formula for~$\tilde{\Sig}_m$ with respect to~$t_0$
and to verify that we obtain zero.
We first observe that taking a solution~$\phi_m \in \H_m$ of the Dirac equation in the presence of~$\B$,
evaluating at time~$t_0$ and applying the time evolution operator~$\tilde{U}^{t,t_0}_m$
gives~$\phi_m$ at time~$t$, i.e.~$\tilde{U}_m^{t,t_0} \phi_m|_{t_0} = \phi_m|_t$.
Differentiating with respect to~$t_0$ yields
\[ \partial_{t_0} \tilde{U}_m^{t,t_0} \phi_m|_{t_0} = 0 \:. \]
The situation is different when one considers the time evolution operator of the vacuum.
Namely, in the expression~$U^{t,t_0}_m \phi_m|_{t_0}$, the wave function~$\phi_m$ satisfies the
Dirac equation~$(i \partial_t - H) \phi_m = \V \phi_m$, whereas the time evolution operator
solves the Dirac equation with~$\V \equiv 0$. As a consequence,
\[ \partial_{t_0} U_m^{t,t_0} \phi_m|_{t_0} =  -i U_m^{t,t_0} (\V \phi_m)|_{t_0} \:. \]
Using these formulas together with~$U^{t_0, t_0} = \1 = \tilde{U}^{t_0, t_0}$,
a straightforward computation gives
\begin{align*}
\partial_{t_0} \big(\psi_m \,|\, \eqref{Sm1} \,\phi_m \big) \big|_{t_0} =& -i (\psi_m \,|\, [\Sig_m, \V]\, \phi_m) |_{t_0} \\
&-\frac{i}{2} \, (-2)
\big( \psi_m \,\big|\, \big( \Sig_m \, \V(t_0) -\V(t_0)\, \Sig_m \big) \,\phi_m \big) \big|_{t_0} \\
&-\frac{i}{2} \int_{-\infty}^\infty \epsilon(t-t_0)\,
\big( (-i \V(t_0)) \,\psi_m \,\big|\, \Sig_m \,U_m^{t_0, t}\, \V(t)\, \tilde{U}_m^{t,t_0} \,\phi_m \big) \big|_{t_0} dt \\
&+\frac{i}{2} \int_{-\infty}^\infty \epsilon(t-t_0)\,
\big( \psi_m \,\big|\, \tilde{U}_m^{t_0, t}\, \V(t)\, \Sig_m \,U_m^{t,t_0} \,(-i \V(t_0)) \,\phi_m \big) \big|_{t_0} dt \\
\partial_{t_0} \big(\psi \,|\, \eqref{Sm2} \,\phi \big)_{t_0} =&
-\frac{1}{2}\int_{-\infty}^\infty \epsilon(t'-t_0) \:
\big( \psi_m \,\big|\, \V(t_0) \,\Sig_m\,U_m^{t_0, t'}\, \V(t')\,
\tilde{U}_m^{t',t_0}\, \phi_m \big) \big|_{t_0} \:dt' \\
&-\frac{1}{2}\int_{-\infty}^\infty \epsilon(t-t_0) \:
\big( \psi_m \,\big|\, \tilde{U}_m^{t_0, t}\, \V(t) \,\Sig_m\,U_m^{t, t_0}\, \V(t_0)\, \phi_m \big) \big|_{t_0} \:dt \:,
\end{align*}
where for notational simplicity we here omitted the restrictions~$|_{t_0}$
for the solutions~$\psi_m$ and~$\phi_m$.
Adding the terms gives zero.
 }} \QEDrem \end{Remark}
The remainder of this section is devoted to the proof of Proposition~\ref{prpSrep}.
Our strategy is to combine the Lippmann-Schwinger equation with estimates in momentum space.
We begin with two technical lemmas.

\begin{Lemma} \label{lemmaC1}
Assume that the external potential~$\B$ satisfies condition~\eqref{L1B}.
For any~$t_0 \in \R$, we denote the characteristic functions in the future respectively
past of this hypersurface~$t=t_0$ by~$\chi_{t_0}^\pm(x)$ (i.e.\ $\chi_{t_0}^\pm(x) = \Theta(\pm(x^0-t_0))$,
where~$\Theta$ is the Heaviside function). Then for any~$\psi_m \in \Cisc(\scrM, S\scrM) \cap \H_m$, the
wave function~$k_m(\chi_{t_0}^\pm \B \psi_m)$ is a well-defined vector in~$\H_{t_0}$ and
\[ \|k_m(\chi_{t_0}^\pm \B \psi_m)\|_{t_0} \leq \frac{1}{2 \pi}\, \|\psi_m\|_m
\int_{-\infty}^\infty \chi_{t_0}^\pm(\tau)\:|\B(\tau)|_{C^0} \:d\tau \:. \]
\end{Lemma}
\Proof Using the integral kernel representation~\eqref{Ukern} and~\eqref{Ukm}
together with the fact that the time evolution in the vacuum is unitary, we obtain
\begin{align*}
2 \pi &\left\| \int_{\R^3} k_m \big( (t_0, .), (\tau, \vec{y}) \big)
\: \big(\chi_{t_0}^\pm \B \psi_m \big)(\tau, \vec{y})\: d^3y \right\|_{t_0} \\
&= \big\| U^{t_0, \tau}_m \gamma^0 (\chi_{t_0}^\pm \B \psi_m)|_\tau \big\|_{t_0}
= \big\|\gamma^0 (\chi_{t_0}^\pm \B \psi_m)|_\tau \big\|_\tau \leq |\B(\tau)|_{C^0} \:\|\psi_m\|_m \:.
\end{align*}
Integrating over~$\tau$ and using~\eqref{L1B} gives the result.
\QED

The following lemma is proved in~\cite[Eqs.~(2.13)--(2.17)]{grotz}.
\begin{Lemma} \label{lemmadistrel}
In the Minkowski vacuum, the fundamental solution~$k_m$ and the Green's function~$s_m$
defined by
\beq \label{smean}
s_m := \frac{1}{2}\: \big( s_m^\vee + s_m^\wedge \big)
\eeq
satisfy the distributional relations in the mass parameters~$m$ and~$m'$
\begin{align*}
k_m\,k_{m'} &= \delta(m-m')\:p_m \\
k_m \,s_{m'}&= s_{m'}\,k_m=\frac{\text{\rm{PP}}}{m-m'}\:k_m \\
s_m\,s_{m'}&= \frac{\text{\rm{PP}}}{m-m'}\:(s_m-s_{m'})+\pi^2 \,\delta(m-m')\:p_m\;,
\end{align*}
where~${\rm{PP}}$ denotes the principal part, and~$p_m$ is the distribution
\beq
\label{pmdef}
p_m(k) = (\slashed{k} + m) \: \delta(k^2 - m^2) \:.
\eeq
\end{Lemma}

\Proof[Proof of Proposition~\ref{prpSrep}]
Let~$\psi \in \H^\infty$ be a family of solutions of the Dirac equation for varying mass.
We denote the boundary values at time~$t_0$ by~$\psi^0_m := \psi_m|_{t_0}$.
Then we can write the Lippmann-Schwinger equation~\eqref{lse} as
\[ \psi_m|_t = U_m^{t,t_0} \psi^0_m +i \int_{t_0}^t U_m^{t, \tau} \big( \gamma^0 \B\: \psi_m \big)
\big|_\tau\: d\tau \:. \]
We now bring this equation into a more useful form.
Expressing the time evolution operator with the help of~\eqref{Ukm} in terms of the fundamental solution,
we obtain
\begin{align*}
\psi_m(x) &= 2 \pi \int_{\R^3} k_m \big( x, (t_0, \vec{y}) \big)\: \gamma^0 \psi_m^0(t_0, \vec{y}) \:d^3y \\
&\quad + 2 \pi i \int_{t_0}^{x^0} dy^0 \int_{\R^3} d^3y \: k_m(x,y)(\B\: \psi_m)(y)\:.
\end{align*}
Applying~\eqref{kmdef} and using that the advanced and retarded Green's functions
are supported in the future and past light cones, respectively,
we can rewrite the last integral
in terms of the advanced and retarded Green's functions,
\[ \psi_m = 2 \pi \,k_m \big( \gamma^0 \delta_{t_0} \psi_m^0 \big)
- s_m^\wedge \big(\chi_{t_0}^+ \B \psi_m \big) - s_m^\vee \big( \chi_{t_0}^- \B \psi_m \big) \:, \]
where~$\delta_{t_0}(x) := \delta(t_0-x^0)$ is the Dirac distribution supported on the
hypersurface~$x^0=t_0$. Next, we express the advanced and retarded Green's functions in terms
of the Green's function~\eqref{smean}: According to~\eqref{kmdef}, we have the relations
\[ s_m=s_m^{\vee}-i\pi k_m=s_m^{\wedge}+i\pi k_m \]
and thus
\beq \label{gdef}
\psi_m = k_m g_m - s_m \B \psi_m \qquad \text{with} \qquad
g_m := 2 \pi \,\gamma^0 \delta_{t_0} \psi_m^0  + i \pi\, \epsilon_{t_0} \B \psi_m \:,
\eeq
where~$\epsilon_{t_0}$ is the step function
\[ \epsilon_{t_0}(x) := \epsilon(x^0-t_0) \]
(and we omitted the brackets in expressions like~$k_m g_m \equiv k_m(g_m)$).
Note that the expression~$k_m g_m$ is well-defined according to Lemma~\ref{lemmaC1}.
We also remark that by applying the operator~$(i \Pdd - m)$ to the distribution~$g_m$ in~\eqref{gdef},
one immediately verifies that~$\psi_m$ indeed satisfies the Dirac equation~$(i \Pdd - m) \psi_m = - \B \psi_m$.

Now we can compute the inner product~$\bra \p \psi | \p \psi \ket$ with the help of
Lemma~\ref{lemmadistrel}. Namely, using~\eqref{gdef},
\begin{align*}
\bra \p \psi | \p \psi \ket \:=& \iint_{I \times I} \bra k_m g_m - s_m \B \psi_m \:|\: k_{m'} g_{m'} - s_{m'} \B \psi_{m'} \ket \:
dm\: dm' \\
=& \int_I \Big( \bra g_m \,|\, p_m g_m \ket + \pi^2\, \bra \B \psi_m \,|\, p_m \B \psi_m \ket \Big) dm \\
&+ \iint_{I \times I} \frac{\text{PP}}{m-m'} \Big(
\bra \B \psi_m \,|\, k_{m'} g_{m'} \ket
- \bra k_m g_m  \,|\, \B \psi_{m'} \ket \\
&\qquad\qquad\qquad\qquad + \bra \B \psi_m \,|\, (s_m - s_{m'}) \B \psi_{m'} \ket \Big) \:dm\: dm' \:.
\end{align*}
Note that this computation is mathematically well-defined in the distributional sense
because~$\psi_m$ and~$g_m$ are smooth and compactly supported
in the mass parameter~$m$.
Employing the explicit formula for~$g_m$ in~\eqref{gdef}, we obtain
\[ \bra \p \psi | \p \psi \ket =
\int_I \Big( \bra g_m \,|\, p_m g_m \ket + \pi^2\, \bra \B \psi_m \,|\, p_m \B \psi_m \ket \Big) dm \:. \]
Comparing~\eqref{kmp} with~\eqref{pmdef} and taking into account that
the operator~$\Sig_m$ defined by~\eqref{S0def} gives a minus sign for the states of negative frequency,
we get
\[ p_m = \Sig_m \,k_m \:. \]
Using this identity together with Proposition~\ref{prpdual} in the vacuum yields the relations
\begin{align*}
\bra g_m \,|\, p_m g_m \ket  &= (k_m g_m \,|\, \Sig_m \,k_m g_m) |_{t_0} \\
\bra \B \psi_m \,|\, p_m \B \psi_m \ket &= ( k_m \B \psi_m \,|\, \Sig_m \,k_m \B \psi_m) |_{t_0}\:.
\end{align*}
We finally apply Proposition~\ref{prp21} to obtain the representation
\beq \label{stiprel}
\bra \p \psi | \p \psi \ket =
\int_I \Big( ( h_m \,|\, \Sig_m \,h_m )|_{t_0} + \pi^2\, ( k_m \B \psi_m \,|\, 
\Sig_m \,k_m \B \psi_m )|_{t_0} \Big) \: dm \:,
\eeq
where
\[ h_m := \psi_m  + i \pi\, k_m ( \epsilon_{t_0} \B \psi_m) \:. \]

Comparing~\eqref{Sdef2} with~\eqref{stiprel}, we get
\[ ( \psi_m \,|\, \tilde{\Sig}_m\, \psi_m)_m = ( h_m \,|\, \Sig_m \,h_m )|_{t_0} + \pi^2\, ( k_m \B \psi_m \,|\, 
\Sig_m \,k_m \B \psi_m )|_{t_0} \:. \]
Expressing the operators~$k_m$ according to~\eqref{Ukm} by the time evolution operator
and writing~$\psi_m$ in terms of the initial data as
\[ \psi_m|_t = \tilde{U}^{t,t_0} \psi|_{t_0} \:, \]
we obtain
\begin{align*}
( \psi_m \,|\, &\tilde{\Sig}_m\, \psi_m)_m  = ( \psi | \Sig_m \psi)|_{t_0} - \frac{i}{2} 
\int_{-\infty}^\infty \epsilon(t-t_0)\, \big( \psi \,\big|\, \Sig_m\, U^{t_0, t} \,\V(t)\, \tilde{U}^{t, t_0} \,
\psi \big) \big|_{t_0}\:dt \\
&+\frac{i}{2} \int_{-\infty}^\infty \epsilon(t-t_0)\: \big(U^{t_0, t} \,\V(t)\, \tilde{U}^{t, t_0}\,
\psi \,\big|\,  \Sig_m\, \psi \big) \big|_{t_0}\:dt \\
&+\frac{1}{4} \iint_{\R \times \R} \epsilon(t-t_0) \,\epsilon(t'-t_0)\:
\big( U^{t_0, t} \,\V(t)\, \tilde{U}^{t, t_0}\, \psi \,\big|\, \Sig_m\, U^{t_0, t'} \,\V(t')\,
\tilde{U}^{t', t_0} \,\psi \big) \big|_{t_0}\:dt\, dt' \\
&+\frac{1}{4} \iint_{\R \times \R} \big( U^{t_0, t} \,\V(t)\, \tilde{U}^{t, t_0}\, \psi \,\big|\, \Sig_m\, U^{t_0, t'} \,\V(t')\,
\tilde{U}^{t', t_0} \,\psi \big) \big|_{t_0}\:dt\, dt' \:.
\end{align*}
Rearranging the terms and polarizing gives the result.
\QED

\Proof[Proof of Theorem~\ref{thmC2}]
Since the time evolution operators are unitary and the operators~$\Sig_m$ have norm one
(see~\eqref{S0def}), the representation~\eqref{Sm1} and~\eqref{Sm2} gives rise to
the following estimate for the $\sup$-norm of~$\tilde{S}_m$,
\[ \big\|\tilde{\Sig}_m \big\| \leq 1 + \int_\R |\V(t)|_{C^0}\:dt 
+\iint_{\R \times \R} |\V(t)|_{C^0}\; |\V(t')|_{C^0} \:dt\, dt' \:. \]
The decay assumption~\eqref{L1B} implies that the $\sup$-norm of~$\tilde{S}_m$ is
bounded uniformly in~$m$. Using this fact in~\eqref{Sdef2} gives the inequality~\eqref{smop},
thereby establishing the strong mass oscillation property.
\QED

We finally remark that the uniqueness statement in Theorem~\ref{thmSrep}
implies that~\eqref{Sm1} and~\eqref{Sm2} yields
an explicit representation of the fermionic signature operator
in the presence of a time-dependent external potential.

\section{Hadamard Form of the Fermionic Projector} 
In this section, we will prove Theorem~\ref{thmHadamard}.
In preparation, we derive so-called frequency splitting estimates which
give control of the ``mixing'' of the positive and negative frequencies in the
solutions of the Dirac equation as caused by the time-dependent external potential
(Theorem~\ref{thmfreqmix}). Based on these estimates, we will complete
the proof of Theorem~\ref{thmHadamard} at the end of Section~\ref{secproofHadamard}.

\subsection{Frequency Mixing Estimates} \label{secfreqmix}
For the following constructions, we again choose the hypersurface~$\scrN := \scrN_{t_0}$ at some given time~$t_0$.
Moreover, we always fix the mass parameter~$m>0$.
Since we are no longer considering families of solutions, for ease in notation we omit
the index~$m$ at the Dirac wave functions, the scalar products and the corresponding norms.
We also identify the solution space~$\H_m$
with the Hilbert space~$\H_{t_0}$ of square integrable wave functions on~$\scrN$.
On~$\H_{t_0}$, we can act with the Hamiltonian~$H$ of the vacuum, and using the above identification,
the operator~$H$ becomes an operator on~$\H_m$ (which clearly depends on the choice of~$t_0)$.

We work with a so-called {\em{frequency splitting}} with respect to the vacuum
dynamics. To this end, we decompose the Hilbert space~$\H_m$ as
\[ \H_m = \H_m^+ \oplus \H_m^- \qquad \text{with} \qquad
\H^\pm = \chi^\pm(H) \H_m \:, \]
where~$\chi^\pm$ are the characteristic functions
\beq \label{chipmdef}
\chi^+ := \chi_{[0, \infty)} \qquad \text{and} \qquad \chi^- := \chi_{(-\infty, 0)} \:.
\eeq
For convenience, we write this decomposition in components and
use a block matrix notation for operators, i.e.
\[ \psi = \begin{pmatrix} \psi^+ \\ \psi^- \end{pmatrix} \qquad \text{and} \qquad
A = \begin{pmatrix} A^+_+ & A^+_- \\ A^-_+ & A^-_- \end{pmatrix} \:, \]
where~$A^s_{s'} = \chi^s(H) A\, \chi^{s'}(H)$ and~$s,s' \in \{\pm\}$.

The representation in Proposition~\ref{prpSrep} 
makes it possible to let the fermionic signature operator~$\tilde{\Sig}_m$ act on the Hilbert space~$\H_m$ (for fixed~$m$).
We decompose this operator with respect to the above frequency splitting,
\[ \tilde{\Sig}_m = \SD + \Delta \tilde{\Sig}\:, \qquad \text{where} \qquad
\SD := \tilde{\Sig}^+_+ + \tilde{\Sig}^-_- \quad \text{and} \quad
\Delta \tilde{\Sig} := \tilde{\Sig}^+_- + \tilde{\Sig}^-_+ \:. \]
Thus the operator~$\SD$ maps positive to positive and negative to negative frequencies.
The operator~$\Delta \tilde{\Sig}$, on the other hand, mixes positive and negative frequencies.
In the next theorem, it is shown under a suitable smallness assumption on~$\B$
that the operators~$\chi^\pm(\tilde{\Sig}_m)$ coincide with the projections~$\chi^\pm(H)$, up to
smooth contributions.
The main task in the proof is to control the ``frequency mixing'' as described by the operator~$\Delta \tilde{\Sig}$.
\begin{Thm} \label{thmfreqmix}
Under the assumptions of Theorem~\ref{thmHadamard},
the operators~$\chi^\pm(\tilde{\Sig}_m)$ have the representations
\beq \label{cD2}
\chi^\pm(\tilde{\Sig}_m) = \chi^\pm(H) + \frac{1}{2 \pi i} 
\ointctrclockwise_{\partial B_{\frac{1}{2}}(\pm 1)}
(\tilde{\Sig}_m-\lambda)^{-1}\, \Delta \tilde{\Sig} \: (\SD-\lambda)^{-1}\:d\lambda \:,
\eeq
where the contour integral is an integral operator with a smooth integral kernel.
\end{Thm} \noindent
Here~$B_{\frac{1}{2}}$ denotes the open ball of radius~$1/2$.
The operator~$(\tilde{\Sig}_m-\lambda)^{-1}$ is also referred to as the {\em{resolvent}} of~$\tilde{\Sig}_m$.

This theorem will be proved in several steps. We begin with a preparatory lemma.

\begin{Lemma} Under the assumptions~\eqref{Bdecay} and~\eqref{Bssmall},
the spectrum of~$\SD$ is located in the set
\beq \label{SDspec}
\sigma(\SD) \subset \bigg[ -\frac{3}{2}, -\frac{1}{2} \bigg] \cup  \bigg[ \frac{1}{2}, \frac{3}{2} \bigg] \:.
\eeq
Moreover,
\beq
\chi^\pm(\SD) = \chi^\pm(H) \label{cD1} \:,
\eeq
and the operators~$\chi^\pm(\tilde{\Sig}_m)$ have the representations~\eqref{cD2}.
\end{Lemma}
\Proof Since the subspaces~$\H^\pm$ are invariant under the action of~$\SD$, our task is to show
that the spectrum of~$\SD|_{\H^\pm}$ is positive and negative, respectively.
This statement would certainly be true if we replaced~$\SD$ by~$\Sig_m$, because
the operator~$\Sig_m$ has the eigenvalues~$\pm 1$ with~$\H^\pm$ as the corresponding eigenspaces.
Estimating the representation in Proposition~\ref{prpSrep} with the Schwarz inequality,
we obtain
\[ \big| (\psi | \SD \phi) - (\psi | \Sig_m \phi) \big|
\leq \left(c + \frac{c^2}{2} \right) \|\psi\|\, \|\phi\| \quad \text{with} \quad
c := \int_{-\infty}^\infty |\B(\tau)|_{C^0}\, d\tau\:. \]
Using the assumption~\eqref{Bssmall}, we conclude that
\[ \big| (\psi | \SD \phi) - (\psi | \Sig_m \phi) \big| < \frac{1}{2}\: 
\|\psi\|\, \|\phi\| \qquad \text{for all~$\psi, \phi \in \H_m$}\:. \]
Standard estimates on the continuity of the spectrum (see for example~\cite[\S IV.3]{kato})
yield that the spectrum of~$\SD$ differs by that of the operator~$\Sig_m$ at most by~$1/2$.
This gives~\eqref{SDspec} and~\eqref{cD1}.

In order to prove the representation~\eqref{cD2}, we take the resolvent identity
\[ (\tilde{\Sig}_m-\lambda)^{-1} = (\SD-\lambda)^{-1} - (\tilde{\Sig}_m-\lambda)^{-1}\,
\Delta \tilde{\Sig} \, (\SD-\lambda)^{-1} \:, \]
form the contour integral and apply~\eqref{cD1}. This gives the result.
\QED

The next lemma relates the smoothness of an integral kernel to
the boundedness of the product of the operator with powers of the vacuum Hamiltonian.
\begin{Lemma} Let~$A \in \Lin(\H_m)$ be an operator which maps smooth functions to
smooth functions and has the property that for all~$p, q \in \N$, the operator product
\beq \label{Hpq}
H^q \,A\, H^p \::\: C^\infty_0(\scrN, S\scrM) \rightarrow C^\infty(\scrN, S\scrM)
\eeq
extends to a bounded linear operator on~$\H_m$.
Then, considering~$A$ as an operator on~$\H_m$, this operator can be represented
as an integral operator with a smooth integral kernel, i.e.
\[ (A \psi)(x) = \int_{\scrN} {\mathcal{A}} \big( x,(t_0, \vec{y}) \big)\, \gamma^0\,
\psi(t_0, \vec{y})\, d^3y \qquad \text{with} \qquad
{\mathcal{A}} \in C^\infty(\scrM \times \scrM)\:. \]
\end{Lemma}
\Proof Since in momentum space, the square of the Hamiltonian takes the form
\[ H \big(\vec{k} \big)^2 = \left( \gamma^0 \big( \vec{\gamma} \vec{k} + m \big) \right)^2 =
\big( -\vec{\gamma} \vec{k} + m \big) \big( \vec{\gamma} \vec{k} + m \big) = |\vec{k}|^2 + m^2 \:, \]
the wave function~$\hat{\psi}$ defined by
\[ \hat{\psi}(\vec{k}) := \frac{1}{|\vec{k}|^2 + m^2}\:e^{i \vec{k} \vec{x}_0}\: \Xi \]
for a constant spinor~$\Xi$ and~$\vec{x}_0 \in \R^3$, satisfies the equation
\[ H^2 \,\psi(\vec{x}) = \delta^3(\vec{x}-\vec{x_0}) \,\Xi \:. \]
Moreover, one verifies immediately that~$\psi \in \H_{t_0}$ is square-integrable.
Using the last equation together with~\eqref{Hpq}, we conclude that
\[ H^q A \big( \delta^3(\vec{x}-\vec{x_0}) \,\Xi \big) = H^q A H^2 \psi \in \H_{t_0} \:. \]
Since~$q$ is arbitrary, it follows that~$A$ has an integral representation in the spatial variables,
\[ (A \phi)(\vec{x}) = \int_{\scrN} {\mathcal{A}}(\vec{x}, \vec{y})\,\gamma^0\, \phi(\vec{y})\: d^3y \qquad \text{with} \qquad {\mathcal{A}} \in C^\infty(\scrN \times \scrN)\:. \]
We now extend this integral kernel to~$\scrM \times \scrM$ by solving the Cauchy problem
in the variables~$x$ and~$y$. This preserves smoothness by the global existence
and regularity results for linear hyperbolic equations, giving the result.
\QED

\begin{Lemma} \label{lemmacommute}
Under the assumptions of Theorem~\ref{thmHadamard},
for all~$p \in \N$ the iterated commutator
\[ \Sig^{(p)} := \underbrace{ \Big[H, \big[H, \ldots ,[H}_{\text{$p$ factors}}, \tilde{\Sig}_m] \cdots \big] \Big] \]
is a bounded operator on~$\H_m$.
\end{Lemma}
\Proof In the vacuum, the Hamiltonian clearly commutes with the time evolution operator,
\beq \label{comm0}
\big[ H, U_m^{t,t'} \big] = 0 \:.
\eeq
In order to derive a corresponding commutator relation in the presence of the external potential,
one must take into account that~$\tilde{H}$ is time-dependent. For ease in notation, we do not
write out this dependence, but instead understand that the Hamiltonian is to be evaluated at the
correct time, i.e.
\[ \tilde{U}_m^{t,t'} \, \tilde{H} \equiv \tilde{U}_m^{t,t'} \, \tilde{H}(t') \qquad \text{and} \qquad
\tilde{H}\, \tilde{U}_m^{t,t'} \equiv \tilde{H}(t)\, \tilde{U}_m^{t,t'} \:. \]
Then
\[ (i \partial_t - \tilde{H}) \big( \tilde{H} \,\tilde{U}_m^{t,t'} - \tilde{U}_m^{t,t'} \tilde{H} \big) = i \dot{\tilde{H}} \,\tilde{U}_m^{t,t'} \qquad \text{and}
\qquad \tilde{H} \,\tilde{U}_m^{t,t'} - \tilde{U}_m^{t,t'} \tilde{H} \big|_{t=t'}=0 \]
(here and in what follows the dot denotes the partial derivative with respect to~$t$).
Solving the corresponding Cauchy problem gives
\beq
\big[ \tilde{H}, \tilde{U}_m^{t,t'} \big] = \int_{t'}^t \tilde{U}_m^{t,\tau} \dot{\tilde{H}} \,\tilde{U}_m^{\tau,t'}\: d\tau
\:. \label{comm1}
\eeq

In order to compute the commutator of~$H$ with the operator products in~\eqref{Sm1}
and~\eqref{Sm2},
we first differentiate the expression~$U_m^{t'',t} \,\V\, \tilde{U}_m^{t,t'}$ with respect to~$t$,
\beq \label{prelim}
i \partial_t \big( U_m^{t'',t} \,\V\, \tilde{U}_m^{t,t'} \big) =
i U_m^{t'',t} \,\dot{\V}\, \tilde{U}_m^{t,t'} + U_m^{t'',t} \,\V\, \tilde{H}\, \tilde{U}_m^{t,t'} - U_m^{t'',t} \,H \,\V\, \tilde{U}_m^{t,t'} \:.
\eeq
Moreover, using the commutation relations~\eqref{comm0} and~\eqref{comm1}, we obtain
\begin{align*}
H &\, (U_m^{t'',t} \,\V\, \tilde{U}_m^{t,t'}) - (U_m^{t'',t} \,\V\, \tilde{U}_m^{t,t'}) \,\tilde{H} \\
&=  U_m^{t'',t}\, H \, \V\, \tilde{U}_m^{t,t'} - U_m^{t'',t}\, \V\, \tilde{H}\, \tilde{U}_m^{t,t'}
+  U_m^{t'',t}\, \V \,[\tilde{H}, \tilde{U}_m^{t,t'}] \\
&= i U_m^{t'',t} \,\dot{\V}\, \tilde{U}_m^{t,t'} - i \partial_t \big( U_m^{t'',t} \,\V\, \tilde{U}_m^{t,t'} \big)
+ \int_{t'}^t  U_m^{t'',t}\, \V\, \tilde{U}_m^{t,\tau} \dot{\tilde{H}} \,\tilde{U}_m^{\tau,t'}\: d\tau \:,
\end{align*}
where in the last step we applied~\eqref{prelim}. It follows that
\begin{align*}
\big[ &H , U_m^{t'',t} \,\V\, \tilde{U}_m^{t,t'} \big] = 
H \, (U_m^{t'',t} \V \tilde{U}_m^{t,t'}) - (U_m^{t'',t} \V \tilde{U}_m^{t,t'}) \,\tilde{H}
+ (U_m^{t'',t} \V \tilde{U}_m^{t,t'}) \,\V \\
&= i U_m^{t'',t} \,\dot{\V}\, \tilde{U}_m^{t,t'} + (U_m^{t'',t} \V \tilde{U}_m^{t,t'}) \,\V
- i \partial_t \big( U_m^{t'',t} \,\V\, \tilde{U}_m^{t,t'} \big)
+ \int_{t'}^t  U_m^{t'',t}\, \V\, \tilde{U}_m^{t,\tau} \dot{\tilde{H}} \,\tilde{U}_m^{\tau,t'}\: d\tau \:.
\end{align*}
Proceeding in this way, one can calculate the commutator of~$H$ with
all the terms in~\eqref{Sm1} and~\eqref{Sm2}. We write the result symbolically as
\[ [H, \tilde{\Sig}_m] = \Sig^{(1)} \:, \]
where~$\Sig^{(1)}$ is a bounded operator.
Higher commutators can be computed inductively, giving the result.
\QED
We point out that this lemma only makes a statement on the iterative commutators.
Expressions like~$[H^p, \tilde{\Sig}_m]$ or~$H^q \,\tilde{\Sig}_m\, H^p$ will not be bounded operators in general.
However, the next lemma shows that the operator~$\Delta \tilde{\Sig}$ has the remarkable property
that multiplying by powers of~$H$ from the left and/or right again gives a bounded operator.
\begin{Lemma} \label{lemmaprod} Under the assumptions of Theorem~\ref{thmHadamard},
for all~$p, q \in \N \cup \{0\}$ the product $H^q \,\Delta \tilde{\Sig}\, H^p$ is a bounded operator on~$\H_m$.
\end{Lemma}
\Proof We only consider the products~$H^q \,\Sig^-_+\, H^p$ because
the operator~$\Sig^+_-$ can be treated similarly.
Multiplying~\eqref{comm1} from the left and right by the resolvent of~$H$, we obtain
\[ \big[(H - \mu)^{-1}, \tilde{\Sig}_m \big] = - (H - \mu)^{-1} \,\Sig^{(1)}\, (H - \mu)^{-1} \:. \]
Writing the result of Lemma~\ref{lemmacommute} as
\[ [H, \Sig^{(p)}] = \Sig^{(p+1)} \qquad \text{with} \qquad \Sig^{(p+1)} \in \Lin(\H) \]
yields more generally the commutation relations
\begin{align}
\big[ (H-\mu)^{-1}, \Sig^{(p)} \big] = - (H - \mu)^{-1} \,\Sig^{(p+1)}\, (H - \mu)^{-1}
\qquad \text{for~$p \in \N$}\:. \label{rescomm}
\end{align}

\begin{figure} %
\begin{picture}(0,0)%
\includegraphics{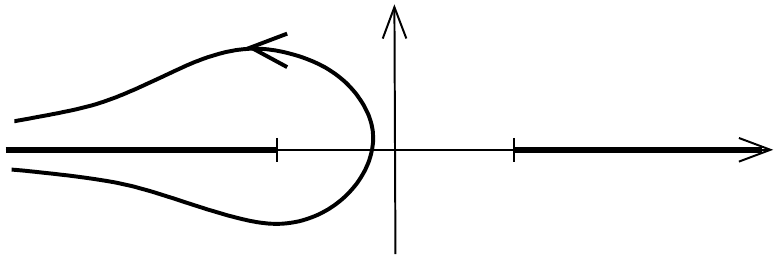}%
\end{picture}%
\setlength{\unitlength}{2486sp}%
\begingroup\makeatletter\ifx\SetFigFont\undefined%
\gdef\SetFigFont#1#2#3#4#5{%
  \reset@font\fontsize{#1}{#2pt}%
  \fontfamily{#3}\fontseries{#4}\fontshape{#5}%
  \selectfont}%
\fi\endgroup%
\begin{picture}(5916,1949)(1037,-6368)
\put(5326,-5401){\makebox(0,0)[lb]{\smash{{\SetFigFont{11}{13.2}{\familydefault}{\mddefault}{\updefault}$\sigma(H)$}}}}
\put(2341,-5401){\makebox(0,0)[lb]{\smash{{\SetFigFont{11}{13.2}{\familydefault}{\mddefault}{\updefault}$\sigma(H)$}}}}
\put(4801,-5866){\makebox(0,0)[lb]{\smash{{\SetFigFont{11}{13.2}{\familydefault}{\mddefault}{\updefault}$m$}}}}
\put(2874,-5873){\makebox(0,0)[lb]{\smash{{\SetFigFont{11}{13.2}{\familydefault}{\mddefault}{\updefault}$-m$}}}}
\put(1262,-5115){\makebox(0,0)[lb]{\smash{{\SetFigFont{11}{13.2}{\familydefault}{\mddefault}{\updefault}$\gamma$}}}}
\end{picture}%

\caption{The contour~$\gamma$.}
\label{figcontour}
\end{figure}

Choosing a contour~$\gamma$ which encloses the interval~$(-\infty, -m]$
as shown in  Figure~\ref{figcontour}, one finds
\begin{align*}
H \Sig^-_+ &= -\frac{1}{2 \pi i} \int_\gamma \mu\, (H - \mu)^{-1} \,\tilde{\Sig}_m \:\chi^+(H)\:d\mu \\
&= \Sig\: H\, \chi^-(H)\: \chi^+(H) + 
\frac{1}{2 \pi i} \int_\gamma \mu\, (H - \mu)^{-1} \,\Sig^{(1)} (H - \mu)^{-1} \:\chi^+(H)\:d\mu \\
&= \frac{1}{2 \pi i} \int_\gamma \mu\, (H - \mu)^{-1} \,\Sig^{(1)} (H - \mu)^{-1} \:\chi^+(H)\:d\mu \:,
\end{align*}
where in the last step we used that~$\chi^-(H)\, \chi^+(H) = 0$.
In order to show that this operator product is bounded,
it is useful to employ the spectral theorem for~$H$,
which we write as
\beq \label{specthm}
f(H) = \int_{\R \setminus [-m,m]} f(\lambda)\, dE_\lambda\:,
\eeq
where~$dE_\lambda$ is the spectral measure of~$H$.
This gives
\begin{align}
H \,\Sig^-_+ &= \iint_{\R \times \R} \bigg(
\frac{1}{2 \pi i} \int_\gamma \frac{\mu}{\lambda - \mu}\: \: \frac{1}{\lambda' - \mu} \:\chi^+(\lambda')
\: dE_\lambda \bigg)\: \Sig^{(1)}\:dE_{\lambda'}\: \:d\mu \nonumber \\
&= -\iint_{\R \times \R} \frac{\lambda}{\lambda - \lambda'}\:
\chi^-(\lambda)\:\chi^+(\lambda') \: dE_\lambda\: \Sig^{(1)}\:dE_{\lambda'}\:. \label{H0S}
\end{align}
Note that the term~$\lambda-\lambda'$ is bounded away from zero.
Thus the factor~$\lambda/(\lambda-\lambda')$ is bounded, showing that the
operator~$H \Sig^-_+$ is in~$\Lin(\H_m)$.

This method can be iterated. To this end, we first rewrite the product with commutators,
\begin{align*}
H^q \,\Sig^-_+ &= \chi^-(H) \:\big( H^- \,\chi^-(H) \big)^p \,\tilde{\Sig}_m\, \chi^+(H) \\
&= \chi^-(H) \Big[ H^-, \big[ H^-, \ldots, [ H^-, \Sig ] \cdots \big] \Big]\:
\chi^+(H) \:,
\end{align*}
where we used the abbreviation~$H^- := H \,\chi^-(H)$.
Multiplying from the right by~$H^p$, we can commute factors~$H^+:= H\, \chi^+(H)$ to the left to obtain
\[ H^q \,\Sig^-_+ H^p = (-1)^p\: \chi^-(H) 
\,\underbrace{\big[ H^+, \ldots, \big[ H^+}_{\text{$p$ factors}} 
, \underbrace{\big[H^-, \ldots, \big[ H^-}_{\text{$q$ factors}}, \tilde{\Sig}_m \big] \cdots \big]
\big] \cdots \big] \, \chi^+(H)\:. \]
Representing each factor~$H^\pm$ by a contour integral, one can
compute the commutators inductively with the help~\eqref{rescomm}.
Applying the spectral theorem~\eqref{specthm} to the left and right of the resulting factor~$\Sig^{(p+q)}$
yields a constant times the expression
\begin{align*} \iint_{\R \times \R} \chi^-(\lambda)\:\chi^+&(\lambda') \: dE_\lambda \,\Sig^{(p+q)}\, dE_{\lambda'} \\
&\times \ointctrclockwise_{\gamma_1} \frac{\mu_1\: d\mu_1}{(\lambda-\mu_1)(\lambda'-\mu_1)} \cdots
\ointctrclockwise_{\gamma_{p+q}} \frac{\mu_{p+q}\: d\mu_{p+q}}{(\lambda-\mu_{p+q})(\lambda'-\mu_{p+q})} \:
\:.
\end{align*}
Carrying out the contour integrals with residues, we obtain similar to~\eqref{H0S}
an expression of the form
\[ H^q \,\Sig^-_+ H^p = \iint_{\R \times \R} f(\lambda, \lambda')\:
\chi^-(\lambda)\:\chi^+(\lambda') \; dE_\lambda\: \Sig^{(p+q)}\:dE_{\lambda'} \]
with a bounded function~$f$. This concludes the proof.
\QED

\Proof[Proof of Theorem~\ref{thmfreqmix}]
It remains to be shown that the contour integral in~\eqref{cD2} has a smooth integral kernel.
To this end, we multiply the integrand from the left by~$H^q$ and from the right by~$H^p$
and commute the factors~$H$ iteratively to the inside.
More precisely, we use the formula
\[ H^q (\tilde{\Sig}_m-\lambda)^{-1} = \sum_{a=0}^q \underbrace{ \big[H, \ldots, [H}_{\text{$a$ factors}},
(\tilde{\Sig}_m-\lambda)^{-1} ] \cdots \big] \:H^{q-a} \]
(note that the sum is telescopic; here we use the convention that the summand for~$a=0$
is simply~$(\tilde{\Sig}_m-\lambda)^{-1} H^q$). Hence
\begin{align*}
&H^q\, (\tilde{\Sig}_m-\lambda)^{-1}\, \Delta \tilde{\Sig} \, (\SD-\lambda)^{-1} \,H^p \\
&= \sum_{a=0}^q \sum_{b=0}^p\underbrace{\Big[H, \ldots, \big[H}_{\text{$a$ factors}},
(\tilde{\Sig}_m-\lambda)^{-1} \big] \cdots \Big] \:H^{q-a} \, \Delta \tilde{\Sig} \, H^{p-b} \: \Big[ \cdots \big[ (\SD-\lambda)^{-1}, \underbrace{H\big], \ldots, H\Big]}_{\text{$b$ factors}} .
\end{align*}
According to Lemma~\ref{lemmaprod}, the intermediate product~$H^{q-a} \, \Delta \tilde{\Sig} \, H^{p-b}$
is a bounded operator. Moreover, the commutators can be computed inductively
with the help of Lemma~\ref{lemmacommute} and the formula
\[ \big[H, (\tilde{\Sig}_m-\lambda^{-1}) \big]
= -(\tilde{\Sig}_m-\lambda^{-1})\,\big[ H, \tilde{\Sig}_m \big]\, (\tilde{\Sig}_m-\lambda^{-1}) \]
(and similarly for~$\SD$). This gives operators which are all bounded
for~$\lambda \in \partial B_\frac{1}{2}(\pm 1)$. Since the integration contour is compact, the
result follows.
\QED

\subsection{Proof of the Hadamard Form} \label{secproofHadamard}
Relying on the frequency mixing estimates of the previous section, we can now
give the proof of Theorem~\ref{thmHadamard}. Recall that the fermionic projector is given by
(see~\eqref{Pdef})
\beq \label{Pshort}
P = -\chi^-(\tilde{\Sig}_m)\, \tilde{k}_m \:,
\eeq
where we again used the short notation~\eqref{chipmdef}.
Here again the operator~$\chi^-(\tilde{\Sig}_m)$ acts on the solution space~$\H_m$ of the
Dirac equation, which can be identified with the space~$\H_{t_0}$ of square integrable
wave functions at time~$t_0$ (see the beginning of Section~\ref{secfreqmix}). 
For the following arguments, it is important to note
that this identification can be made at any time~$t_0$.

In order to prove that the bi-distribution corresponding to~$P$
is of Hadamard form, we compare the fermionic projectors
for three different Dirac operators
and use the theorem on the propagation of singularities
in~\cite{sahlmann2001microlocal}.
More precisely, we consider the following three fermionic projectors:
\begin{enumerate}
\item The fermionic projector~$P^\text{vac}$ in the Minkowski vacuum. \\[-0.85em]
\item The fermionic projector~$\breve{P}$ in the presence of the external potential
\[ \breve{\B}(x) := \eta \big(x^0 \big)\: \B(x) \:, \]
where~$\eta \geq 0$ is a smooth function with~$\eta|_{(-\infty, 0)} \equiv 0$
and~$\eta|_{(1, \infty)} \equiv 1$. \\[-0.85em]
\item The fermionic projector~$P$ in the presence of the external potential~$\B(x)$.
\end{enumerate}
The potential~$\breve{\B}$ vanishes for negative times, whereas for times~$x^0>1$
it coincides with~$\B$. Thus it smoothly interpolates between the dynamics
with and without external potential. The specific form of the potential~$\breve{\B}$
in the transition region~$0 \leq x^0 \leq 1$ is of no relevance for our arguments.

In the Minkowski vacuum, the relation~\eqref{Pshort} gives the usual two-point function
composed of all negative-frequency solutions of the Dirac equation.
It is therefore obvious that the bi-distribution~$P^\text{vac}(x,y)$ is of Hadamard form.

We now compare~$P^\text{vac}$ with~$\breve{P}$. To this end, we choose an
arbitrary time~$t_0<0$. Then, applying the result of Theorem~\ref{thmfreqmix} to~\eqref{Pshort}, we get
\[ P^\text{vac} = -\chi^-(H)\, k_m \qquad \text{and} \qquad
\breve{P} = -\chi^-(H)\, \breve{k}_m + \text{(smooth)} \:, \]
where~$\breve{k}_m$ is the causal fundamental solution in the presence of the potential~$\breve{\B}$.
Since~$\breve{\B}$ vanishes in a neighborhood of the Cauchy surface at time~$t_0$,
we conclude that~$P^\text{vac}$ and~$\breve{P}$ coincide in this neighborhood up to a smooth contribution.
It follows that also~$\breve{P}(x,y)$ is of Hadamard form in this neighborhood.
Using the theorem on the propagation of singularities~\cite[Theorem 5.5]{sahlmann2001microlocal},
we conclude that~$\breve{P}(x,y)$ is of Hadamard form for all~$x,y \in \scrM$.

Next, we compare~$\breve{P}$ with~$P$. Thus we choose an arbitrary time~$t_0>1$.
Using again the result of Theorem~\ref{thmfreqmix} in~\eqref{Pshort}, we obtain
\[ \breve{P} = -\chi^-(H)\, \breve{k}_m + \text{(smooth)} \qquad \text{and} \qquad 
P = -\chi^-(H)\, \tilde{k}_m + \text{(smooth)} \]
(where the smooth contributions may of course be different).
Since~$\breve{\B}$ and~$\B$ coincide in a neighborhood of the Cauchy surface at time~$t_0$,
we infer that~$\breve{P}$ and~$P$ coincide in this neighborhood up to a smooth contribution.
As a consequence, $P(x,y)$ is of Hadamard form in this neighborhood.
Again applying~\cite[Theorem 5.5]{sahlmann2001microlocal},
it follows that~$P(x,y)$ is of Hadamard form for all~$x,y \in \scrM$.
This concludes the proof of Theorem~\ref{thmHadamard}.

\section{Quantum Fields and the Hadamard State} \label{secphysics}
In preparation of defining the CAR algebra, we denote the co-spinor bundle
by~$S^*\scrM$ (thus the fiber~$S_x^*\scrM$ is the dual space of~$S_x\scrM$).
On the smooth and compactly supported sections of~$S^*\scrM$, we introduce the dual of
the Dirac operator~$\Dir^*$ by
\begin{align*}
\Dir^* : C^\infty_0(\scrM, S^*\scrM) &\rightarrow C^\infty_0(\scrM, S^*\scrM) \:, \\
\int_\scrM \big((\Dir^*g)(f) \big)(x)\: d^4x &= \int_\scrM \big(g(\Dir f) \big)(x)\: d^4x \qquad
\text{for all~$f \in C^\infty_0(\scrM, S\scrM)$}\:.
\end{align*}
Moreover, we define the space of pairs of spinorial test functions by
\[ \frakD := C^\infty_0(\scrM, S\scrM) \oplus C^\infty_0(\scrM, S^*\scrM) \]
with the topology induced by the family of seminorms
\begin{equation*}
|(f, g) |_{C^k} := \sup_{x \in \scrM} |\partial^k f(x)| + \sup_{y \in \scrM} |\partial^k g(y)|
\end{equation*}
(and~$|.|$ is any norm on the spinors and co-spinors, respectively).
Next, we define the anti-linear involution map
\beq \label{Gammadef}
\Gamma: \frakD \rightarrow \frakD\:, \qquad
\Gamma (f \oplus g) := g^* \oplus f^* \:,
\eeq
where the star again denotes the adjoint with respect to the spin scalar product, i.e.
\begin{align*}
^* &\::\: \;C^\infty_0(\scrM,S\scrM) \,\rightarrow C^\infty_0(\scrM,S^*\scrM), &\hspace*{-1cm}
f^*(\phi) &= \Sl f | \phi \Sr \\
^* &\::\: C^\infty_0(\scrM,S^*\scrM) \rightarrow \;C^\infty_0(\scrM,S\scrM), &\hspace*{-1cm}
\Sl g^* | \phi \Sr &= g(\phi)
\end{align*}
(we remark that the adjoint spinor~$f^*$ can be identified with the adjoint spinor
usually written as~$\overline{f}= f^\dagger \gamma^0$).
Finally, we introduce an inner product on~$\frakD$,
\beq \label{scalDdef}
(.|.)_\frakD \::\: \frakD \times \frakD \rightarrow \C \:, \qquad
\big( f \oplus g \,\big|\, a \oplus b \big)_\frakD = \bra f \,|\, \tilde{k}_m a \ket + \bra b^* \,|\, \tilde{k}_m\, g^* \ket \:,
\eeq
where~$\bra .|. \ket$ is again the space-time inner product~\eqref{stip}
(this notation is consistent with our overall convention that brackets like~$(.|.)$, 
$(.|.)_m$ and~$\bra .|. \ket$ are
{\em{sesquilinear}} forms, meaning that the first argument always involves complex conjugation).
Applying Proposition~\ref{prpdual}, we can write this inner product as
\[ \big( f \oplus g \,\big|\, a \oplus b \big)_\frakD = ( \tilde{k}_m f \,|\, \tilde{k}_m a )_m 
+ ( \tilde{k}_m b^* \,|\, \tilde{k}_m\, g^* )_m \:, \]
showing that it is indeed positive definite. 
Forming the completion, we thus obtain the Hilbert space~$(\H_\frakD, (.|.)_\frakD)$.
A short computation reveals that the involution~$\Gamma$
preserves the scalar product~$(.|.)_\frakD$ and is therefore anti-unitary, i.e.
\beq \label{antiunitary}
\Gamma^2 = \1 \qquad \text{and} \qquad
(\Gamma h \,|\, \Gamma \tilde{h})_\frakD = (\tilde{h} | h)_\frakD \quad \text{for all~$h, \tilde{h}
\in \H_\frakD$}\:.
\eeq

\begin{Def} \label{CAR} The {\bf{field algebra}}~$\mathcal{F}$
is the unital $*$-algebra generated by the abstract element $B(h)$ with $h \in \frakD$,
together with the following relations for all~$f \oplus g \in \frakD$ and~$\alpha, \beta \in \C$:
\begin{itemize}
\item[(i)] Linearity: $B(\alpha f  \oplus g + m \oplus \beta n) = \alpha B( f  \oplus g) + \beta B (m \oplus  n)$ \\[-0.85em]
\item[(ii)] Hermiticity: $B(f\oplus g)^*=B \big( \Gamma (f\oplus g) \big)$ \\[-0.85em]
\item[(iii)] Dynamics: $B\big((\Dir-m) f \oplus (\Dir^*-m) g \big) = 0$ for all~$f \oplus g \in \frakD$ \\[-0.85em]
\item[(iv)] Canonical anti-commutation relations (CARs)
\[ \big\{ B(f\oplus g)^* , B (m \oplus  n) \big\} = (f\oplus g \,|\, m \oplus n)_\frakD \cdot 1_{\mathcal{F}} \:. \]
\end{itemize}
\end{Def} \noindent
We define smeared field operators by
\[ \Psi(g) := B(0 \oplus g) \qquad \text{and} \qquad
\Psi^*(f) := B(f \oplus 0) \:, \]
Then these field operators satisfy
\begin{align*}
\{\Psi(g),\Psi^*(f)\} &= \big\{ B(0 \oplus g),  B(f \oplus 0) \big\}
\overset{\rm{(ii)}}{=} \big\{ B(0 \oplus g),  B \big( \Gamma(f \oplus 0) \big)^* \big\} \\
&\!\overset{\rm{(iv)}}{=} \big( \Gamma(f \oplus 0) \,\big|\,0 \oplus g \big)_\frakD
\overset{\eqref{Gammadef}}{=} \big( 0 \oplus f^* \,\big|\, 0 \oplus g \big)_\frakD
\overset{\eqref{scalDdef}}{=} \bra g^* \,|\, \tilde{k}_m\, f \ket \:,
\end{align*}
giving rise to the usual anti-commutation relations~\eqref{antismeared}.

The next step is to show that the fermionic projector induces a pure, quasi-free state on the field algebra.
Let us recall a few basics. By definition, a {\em{state}}~$\omega$ is a linear functional on $\mathcal{F}$ which is positive
and normalized, i.e.
\[ \omega(B^*B) \geq 0 \;\;\text{ for all~$B\in\mathcal{F}$} \qquad \text{and} \qquad
\omega(1_{\mathcal{F}}) = 1 \:. \]
By linearity, it suffices to specify~$\omega$
on the monomials. This gives rise to the $n$-point distributions~$\omega_n \in (\frakD^n)'$ defined by
$$ \omega_n(h_1,\ldots,h_n) := \omega \big( B(h_1)\cdots B(h_n)\big) \:,$$
also referred to as the $n$-{\em{point functions}}.
A state induces the usual Fock representation of the field algebra:
\begin{Thm} {\bf{(GNS construction)}} \label{GNS}
Let $\omega$ be a state on a unital $*$-algebra $\mathcal{F}$. 
Then there exists a Hilbert space $(\H_\text{\tiny{Fock}},(.\,,.)_\text{\tiny{Fock}})$ 
and a dense subspace $\mathcal{W} \subset \H_\text{\tiny{Fock}}$ as well as
a representation $\pi \::\: {\mathcal{F}} \to \Lin(\mathcal{W})$ and a unit vector $\Omega\in\mathcal{W}$
such that $\omega=(\Omega,\pi(.)\Omega)_\text{\tiny{Fock}}$ and $\mathcal{W}=\pi(\mathcal{F})\Omega$.
The {\em GNS triple} $(\mathcal{W},\pi,\Omega)$ is determined up to unitary equivalence.
\end{Thm} \noindent
Among all possible states, a distinguished role is played by the quasi-free states, for which
the $n$-point functions are all determined by the two-point functions:
\begin{Def} 
A state $\omega: {\mathcal{F}} \rightarrow\C$ is called {\bf quasi-free} if the $n$-point functions
vanish for odd $n$, while for even $n$, one has
\[ \omega_n( h_1, \ldots, h_n)=\sum\limits_{\sigma \in S'_n} (-1)^{\text{\rm{sign}}(\sigma)} \prod\limits_{i=1}^{n/2}
\;\omega_2 \big( h_{\sigma(2i-1)}, h_{\sigma(2i)} \big) \:, \]
where~$S'_n$ denotes the set of ordered permutations of $n$ elements, i.e.
\begin{align*}
\sigma(2i-1) &< \sigma(2i) &&\hspace*{-1cm} \text{for all~$i=1, \ldots, \frac{n}{2}$} \\[0.15em]
\sigma(2i-1) &< \sigma(2i+1) &&\hspace*{-1cm} \text{for all~$i=1, \ldots, \frac{n-2}{2}$}\:.
\end{align*}
\end{Def} \noindent
For the construction of a quasi-free state from the fermionic projector, we rely on the following result due to
Araki~\cite[Lemmas~3.2 and~3.3.]{araki1970quasifree}:
\begin{Lemma}\label{lemmaAraki}
Let $R$ be a bounded symmetric operator on $(\H_\frakD, (.|.)_\frakD)$ with the following properties
\begin{itemize}
\item[(a)] $R+\Gamma R \Gamma = \1$, \\[-0.85em]
\item[(b)] $0\leq R=R^*\leq \1$ .
\end{itemize}
Then there exists a unique quasi-free state~$\omega$ on $\mathcal{F}$ such that
\beq \label{omegaprop}
\omega\big(B(h)^* B(\tilde{h})\big)=(h \,|\, R \,\tilde{h})_\frakD \qquad \text{for all~$h, \tilde{h} \in \H_\frakD$}\:.
\eeq
\end{Lemma} \noindent
Moreover, Araki proves that if~$R$ is a projection operator,
then the state~$\omega$ is pure (see~\cite[Lemma~4.3]{araki1970quasifree}).

Thus our task is to construct an operator~$R$ with the above properties using the
fermionic signature operator.
As will become clear below, the correct choice is to define~$R$ implicitly by
\beq \label{Rdef}
\big( f \oplus g \,|\, R\, (a \oplus b) \big)_\frakD =
\big( \tilde{k}_m f \,|\, \chi^-(\tilde{\Sig}_m)\, \tilde{k}_m a \big)_m
+ \big( \tilde{k}_m b^* \,|\, \chi^+(\tilde{\Sig}_m)\, \tilde{k}_m g^* \big)_m \:,
\eeq
giving a bounded and symmetric operator. Moreover,
since~$\chi^\pm(\tilde{\Sig}_m)$ are projection operators, we know that
\[ 0 \leq \big( f \oplus g \,|\, R\, (f \oplus g) \big)_\frakD
\leq \big( \tilde{k}_m f \,|\, \tilde{k}_m f \big)_m
+ \big( \tilde{k}_m g^* \,|\, \tilde{k}_m g^* \big)_m = \big( f \oplus g \,|\, f \oplus g \big)_\frakD \:, \]
showing that the condition~(b) in Lemma~\ref{lemmaAraki} holds. Next, using that~$\Gamma$
is an anti-unitary involution~\eqref{antiunitary} and~$R$ is symmetric, it follows that
\[ \big( f \oplus g \,|\, \Gamma R \Gamma\, (a \oplus b) \big)_\frakD
= \big( R \Gamma\,  (a \oplus b) \,|\, \Gamma\, (f \oplus g) \big)_\frakD
= \big( \Gamma\,  (a \oplus b) \,|\, R \Gamma \,(f \oplus g) \big)_\frakD \:. \]
Now we can apply~\eqref{Gammadef} and~\eqref{Rdef} to obtain
\begin{align}
\big( f \oplus g \,|\, \Gamma R \Gamma\, (a \oplus b) \big)_\frakD
&= \big( b^* \oplus a^* \,|\, R \, (g^* \oplus f^*) \big)_\frakD \notag \\
&= \big( \tilde{k}_m b^* \,|\, \chi^-(\tilde{\Sig}_m)\, \tilde{k}_m g^* \big)_m
+ \big( \tilde{k}_m f \,|\, \chi^+(\tilde{\Sig}_m)\, \tilde{k}_m a \big)_m \:. \label{GRGdef}
\end{align}
Adding~\eqref{Rdef} and~\eqref{GRGdef}, we get
\[ \big( f \oplus g \,|\, \big( R + \Gamma R \Gamma \big) \, (a \oplus b) \big)_\frakD =
\big( \tilde{k}_m f \,|\, \tilde{k}_m a \big)_m
+ \big( \tilde{k}_m b^* \,|\, \tilde{k}_m g^* \big)_m = 
\big( f \oplus g \,|\, a \oplus b \big)_\frakD \:, \]
proving that the condition~(a) in Lemma~\ref{lemmaAraki} is satisfied.
Thus Lemma~\ref{lemmaAraki} applies, giving a quasi-free state~$\omega$
with the property~\eqref{omegaprop}. Moreover, as is verified by direct computation,
$R$ is a projection operator. Therefore, the state~$\omega$ is pure.

We finally calculate the two-point function. Beginning with the computation
\begin{align*}
\omega \big( \Psi(g)\, \Psi^*(f) \big) &=
\omega\big( B(0 \oplus g)\, B(f \oplus 0) \big)
\overset{(\star)}{=} \omega\Big( B \big(\Gamma(0 \oplus g) \big)^*\, B(f \oplus 0) \Big) \\
&\!\!\overset{\eqref{Gammadef}}{=} \omega\big( B(g^* \oplus 0))^*\, B(f \oplus 0) \big)
\overset{\eqref{omegaprop}}{=} \big( g^* \oplus 0 \,|\, R\, (f \oplus 0) \big)_\frakD \\
&\!\!\overset{\eqref{Rdef}}{=} \big( \tilde{k}_m g^* \,|\, \chi^-(\tilde{\Sig}_m)\, \tilde{k}_m f \big)_m
\overset{\eqref{dual}}{=} \bra g^* \,|\, \chi^-(\tilde{\Sig}_m)\, \tilde{k}_m f \ket
\end{align*}
(where in~$(\star)$ we used the property~(ii) in Definition~\ref{CAR}),
we can apply the definition of the fermionic projector~\eqref{Pdeftilde} to obtain
\[ \omega \big( \Psi(g)\, \Psi^*(f) \big) = -\bra g^* | P f \ket
= -\iint_{\scrM \times \scrM} g(x) P(x,y) f(y) \: d^4x\, d^4y \:. \]
This concludes the proof of Theorem~\ref{thmstate}.

\appendix
\section{Uniform $L^2$-Estimates of Derivatives of Dirac Solutions} \label{appB}
We now derive a few estimates which will be needed for the proof of the mass oscillation
property in Section~\ref{secproofmass}. We use standard methods of the theory of
partial differential equations and adapt them to the Dirac equation.
In generalization of~\eqref{sobolev}, we denote the spatial Sobolev norms by
\[ \| \phi \|^2_{W^{a, 2}} = \sum_{\text{$\alpha$ with~$|\alpha| \leq a$}}\:
\int_{\R^3} | \nabla^\alpha \phi(\vec{x}) |^2\: d^3x \:. \]

\begin{Lemma} \label{lemmaB} We are given two non-negative integers~$a$ and~$b$ as well as
a smooth time-dependent potential~$\B$. In the case~$a>0$ and~$b \geq 0$, we assume furthermore
that the spatial derivatives of~$\B$ decay faster than linearly for large times in the sense that
\beq \label{Bdecay2}
|\nabla \B(t)|_{C^{a-1}} \leq \frac{c}{1 + |t|^{1+\varepsilon}}
\eeq
for suitable constants~$c, \varepsilon>0$. Then there is a constant~$C=C(c, \varepsilon, a,b)$ such
that every family of solutions~$\psi \in \H^\infty$
of the Dirac equation~\eqref{Dout} for varying mass parameter can be estimated for all times in terms of the
boundary values at~$t=0$ by
\[ \big\| \partial^b_m \psi_m|_t \big\|_{W^{a, 2}} \leq C\, \big( 1+|t|^b \big)\: \sum_{p=0}^b \big\| \partial^p_m
\psi_m|_{t=0} \big\|_{W^{a, 2}} \:. \]
\end{Lemma}
\Proof We choose a multi-index~$\alpha$ of length~$a:=|\alpha|$
and a non-negative integer~$b$.
Differentiating the Dirac equation~\eqref{Dout} 
with respect to the mass parameter and to the spatial variables gives
\[ (i \Pdd + \B - m) \:\nabla^\alpha \partial_m^b \psi_m = b \,\nabla^\alpha \partial_m^{b-1}
\psi_m - \nabla^\alpha \big(\B \,\partial_m^b \psi_m \big)
+ \B \,\nabla^\alpha \partial_m^b \psi_m \:. \]
Introducing the abbreviations
\[ \Xi := \nabla^\alpha \partial_m^b \psi_m \qquad \text{and} \qquad
\phi := b\, \nabla^\alpha \partial_m^{b-1} \psi_m - \nabla^\alpha \big(\B \,\partial_m^b \psi_m \big)
+ \B \,\nabla^\alpha \partial_m^b \psi_m \:, \]
we rewrite this equation as the inhomogeneous Dirac equation
\[ (\Dir - m) \,\Xi = \phi \:. \]
A calculation similar to current conservation yields
\[ -i\partial_j \Sl \Xi | \gamma^j \Xi \Sr = \Sl (\Dir -m) \Xi \,|\, \Xi \Sr - \Sl \Xi \,|\, (\Dir -m) \Xi \Sr
= \Sl \phi | \Xi \Sr - \Sl \Xi | \phi \Sr \:. \]
Integrating over the equal time hypersurfaces and using the Schwarz inequality, we obtain
\[ \big| \partial_t \big( \Xi|_t \big| \Xi|_t \big)_t \big|
\leq 2\, \big\|\Xi|_t \big\|_t\: \big\|\phi|_t \big\|_t \]
and thus
\[ \Big| \partial_t \big\| \Xi|_t \big\| \Big| \leq \big\|\phi|_t \big\|_t \:. \]
Substituting the specific forms of~$\Xi$ and~$\phi$ and using the Schwarz and triangle
inequalities, we obtain the estimate
\beq \label{fundes}
\Big| \partial_t \big\| \nabla^\alpha \partial_m^b \psi_m|_t \big\|_t \Big|
\leq b\, \big\| \nabla^\alpha  \partial_m^{b-1} \psi_m|_t \big\|_t + c\, a\,|\nabla \B(t)|_{C^{a-1}}\:
\big\| \nabla^\alpha \partial_m^b \psi_m|_t\big\|_{W^{a-1, 2}}\:,
\eeq
where we used the notation~\eqref{defCk}.

We now proceed inductively in the maximal total order~$a+b$ of the derivatives.
In the case~$a=b=0$, the claim follows immediately from the unitarity of the time evolution.
In order to prove the induction step, we note that in~\eqref{fundes}, the
order of differentiation of the wave function on the right hand side
is smaller than that on the left hand side at least by one.
In the case~$a=0$ and~$b \geq 0$, the induction hypothesis yields the inequality
\[ \big| \partial_t \| \partial_m^b \psi_m|_t \| \big| \leq b\, \big\| \partial_m^{b-1} \psi_m|_t \big\| 
\leq b \,C\: \big(1+|t|^{b-1} \big)\: \sum_{p=0}^{b-1} \big\| \partial^p_m \psi_m|_{t=0} \big\| \:, \]
and integrating this inequality from~$0$ to~$t$ gives the result.
In the case~$a>0$ and~$b \geq 0$, we apply~\eqref{Bdecay2}
together with the induction hypothesis to obtain
\begin{align*}
\Big| \partial_t \big\| \partial_m^b \psi_m|_t \big\| \Big|_{W^{a,2}}
\leq\;& b \,C\: \big(1+|t|^{b-1} \big)\: \sum_{p=0}^{b-1} \big\| \partial^p_m \psi_m|_{t=0} \big\|_{W^{a,2}} \\
&+ c\,C\: \frac{1+|t|^b}{1+|t|^{1+\varepsilon}}\:
\sum_{p=0}^b \big\| \partial_m^p \psi_m|_{t=0} \big\|_{W^{a-1, 2}}\:.
\end{align*}
Again integrating over~$t$ gives the result.
\QED

\Thanks {{\em{Acknowledgments:}}
We would like to thank Claudio Dappiaggi, Chris Fewster, Christian G{\'e}rard 
and Nicola Pinamonti for helpful discussions.
We are grateful to Johannes Kleiner and the referee for useful comments on the manuscript.

\providecommand{\bysame}{\leavevmode\hbox to3em{\hrulefill}\thinspace}
\providecommand{\MR}{\relax\ifhmode\unskip\space\fi MR }
\providecommand{\MRhref}[2]{%
  \href{http://www.ams.org/mathscinet-getitem?mr=#1}{#2}
}
\providecommand{\href}[2]{#2}

\end{document}